\documentclass[10pt,aps,pra,twocolumn]{revtex4-2} 

\usepackage{amsmath}

\newcommand{\bear}{\begin{eqnarray}}
\newcommand{\eear}{\end{eqnarray}}
\newcommand{\be}{\begin{equation}}
\newcommand{\ee}{\end{equation}}
\newcommand{\beqn}{\begin{eqnarray}}
\newcommand{\eeqn}{\end{eqnarray}}
\newcommand{\beqnn}{\begin{eqnarray*}}
\newcommand{\eeqnn}{\end{eqnarray*}}

\def\vf{\varphi}
\def\vep{\varepsilon}

\begin{document}

\title{Uncertainty relations: a small zoo of remarkable inequalities discovered since 1927
}

\author{V. V. Dodonov
(orcid {0000-0001-7599-209X})
}

\email{vicdod@gmail.com}

\affiliation{Universidade de Brasília, Centro Internacional de Física, 70297-400, Brasília,  DF, Brasil.}

\begin{abstract}
A concise review of various mathematical formulations
of the uncertainty relations in quantum mechanics discovered since 1927 is given.
Besides the traditional Heisenberg inequality,
the modifications made by Schr\"odinger and Robertson, as well
as generalizations to sets of several noncommuting operators,
are considered. The "entropic" inequalities and "local"
uncertainty relations, together with inequalities which connect
the so-called total width and the mean peak width of a wave function,
are discussed. Inequalities for the products of higher order moments
of the coordinate and momentum are presented. Inequalities
making the uncertainty relations more accurate when the "purity"
of a quantum state is fixed are demonstrated. Diverse formulations of
the energy-time uncertainty relations are considered.

 
\end{abstract}

\maketitle   

\section{Introduction}

The famous ``uncertainty relation'' (UR)
\be
\Delta x \Delta p \ge \hbar/2
\label{dxp}
\ee
was introduced  by Heisenberg  \cite{Heis} in 1927 as an approximate (qualitative) inequality. 
Later in the same year, 
one of its quantitative formulations (actually, the simplest possible one) was proven 
rigorously in the frameworks of the wave function description of
quantum systems by Kennard \cite{Kennard}. Since that time, relation (\ref{dxp})
is frequently considered as one of cornerstones of quantum mechanics. {For example, Feynman told in his lectures 
\cite{Feynman}: ``The uncertainty principle ``protects'' quantum mechanics. Heisenberg recognized that if it were 
possible to measure the momentum and the position simultaneously with a greater accuracy, 
the quantum mechanics would collapse.'' }

Since this subject is discussed  in every textbook on quantum mechanics,
one could imagine that it was closed many decades ago.
Nonetheless, it is remarkable (and, perhaps, surprising) that new papers, devoted to generalizations of the UR and their consequences,
still appear in the physical and mathematical literature.
Moreover, some ``burst'' of publications on this subject is observed during a few past years, related mainly to problems of the quantum information theory.

The aim of this mini-review is to bring examples of the most impressive results achieved during almost
100 years passed since 1927. In the most cases, only the final results will be given.
Detailed proofs can be found in numerous references.


\section{Uncertainty Relations Based on Variances}

Following Heisenberg \cite{Heis}, the "uncertainty" of a quantity A is frequently
defined as the square root of its variance (or mean squared deviation)
$\sigma_A \equiv \langle \hat{A}^2\rangle - \langle \hat{A}\rangle^2$, i.e.,
$\Delta A \equiv \sqrt{\sigma_A}$.
Here $\hat{A}$ is the Hermitian operator, corresponding to the observable $A$, 
and the angular brackets mean averaging over the state
of the quantum system:
$\langle \hat{A}\rangle = \int \psi^* \hat{A}\psi\,dV$
for a pure state described by means of the normalized wave function $\psi$
or $\langle \hat{A}\rangle = \mbox{Tr}(\hat\rho\hat{A})$
for a mixed state described by means of the Hermitian positive definite operator $\hat\rho$
with
$\mbox{Tr}(\hat\rho) =1$.
Note that Heisenberg considered only Gaussian wave packets, for
which relation (1) is an equality, whereas in more general
situations he confined himself to the analysis of several ``thought
experiments'' 
(considered in a more detailed form by Bohr \cite{Bohr28}). 
For arbitrary pure states, the inequality
\be
\sigma_p \sigma_x \ge \hbar^2/4
\label{unc2}
\ee
was obtained for the first
time by Kennard \cite{Kennard}. 

 A generalization of relation (\ref{unc2}) to the case of ``classical''
observables $A$ and $B$, i.e., some functions of the canonically
conjugate coordinates and momenta, 
was made by Robertson \cite{Robertson29} (for pure quantum states):
\be
\sigma_A \sigma_B \ge \frac14\left|
\langle [\hat{A},\hat{B}]\rangle\right|^2.
\label{unc3}
\ee
However, simple examples show that in many cases the left-hand side of (\ref{unc3}) turns out 
bigger than the right-hand one. This means that, probably, one should add some extra terms to
the right-hand side, taking into account some additional parameters or specific properties
of concrete quantum systems under consideration.
The first step in this direction was made by  Schr\"odinger in paper \cite{Schrod30} 
(its translation into English can be found, e.g., in paper \cite{Angelow99}).
He  obtained a more precise version of (\ref{unc3}),
taking into account the average value of the anticommutator of operators $\hat{A}$ and $\hat{B}$:
\be
\sigma_A \sigma_B \ge \sigma_{AB}^2 + \frac14\left|
\left\langle [\hat{A},\hat{B}]\right\rangle\right|^2 
\equiv G_{AB}^2,
\label{unc4}
\ee
\be
\sigma_{AB} \equiv \frac12 \left\langle \Big\{\delta\hat{A} , \delta\hat{B} \Big\}\right\rangle, 
\qquad \delta\hat{A} \equiv \hat{A} -\langle \hat{A}\rangle.
\label{defsigAB}
\ee
The same inequality (\ref{unc4}) was given by Robertson \cite{Robertson30}, but he wrote the right-hand side 
in a different form, which is equivalent to 
$G_{AB}^2 \equiv \left|\left\langle (\delta\hat{A})(\delta\hat{B})\right\rangle\right|^2$.

Applying (\ref{unc4}) to the coordinate and momentum operators, one arrives at
the following generalization of (\ref{unc2}): 
\be
\sigma_p \sigma_x -\sigma_{xp}^2 \ge \hbar^2/4 .
\label{unc5}
\ee
Note, however, that inequality (\ref{unc5}) was not written explicitly in Refs. \cite{Schrod30,Robertson30}.
Inequality (\ref{unc5}) can be rewritten in the form \cite{DKM80,183}
\be
\sigma_p \sigma_x \ge \frac{\hbar^2}{4\left(1-r^2\right)}, \qquad
r= \frac{\sigma_{xp}}{\sqrt{\sigma_p \sigma_x}},
\label{unc34}
\ee
demonstrating the role of the ``correlation coefficient'' $r$ as an additional parameter 
responsible for the increase of product $\sigma_p \sigma_x$. 

In the case of two
arbitrary (not necessarily Hermitian) operators $\hat{A}$ and $\hat{B}$, 
the main inequality has the form \cite{DKM80}
\beqn
\left\langle \delta\hat{A}^{\dagger}\delta\hat{A}\right\rangle
\left\langle \delta\hat{B}^{\dagger}\delta\hat{B}\right\rangle
&\ge& \frac14 \left|\left\langle \delta\hat{A}^{\dagger}\delta\hat{B}
- \delta\hat{B}^{\dagger}\delta\hat{A}\right\rangle\right|^2
\nonumber \\ &&
+ \frac14 \left\langle \delta\hat{A}^{\dagger}\delta\hat{B}
+ \delta\hat{B}^{\dagger}\delta\hat{A}\right\rangle^2.
\label{unc14}
\eeqn
 A simple example of non-Hermitian operators is  $\hat{A} = \hat{a}$
and $\hat{B} = \hat{a}^{\dagger}$,
where $\hat{a}$ and $\hat{a}^{\dagger}$ are
the boson annihilation and creation operators: $[\hat{a}, \hat{a}^{\dagger}] = 1$.
 Then, the following inequality must hold
for the quantities 
$\sigma_a= \left\langle \hat{a}^2\right\rangle - \left\langle \hat{a}\right\rangle^2$
and
$\delta_N= \left\langle \hat{a}^{\dagger}\hat{a}\right\rangle
- \left|\left\langle \hat{a}\right\rangle\right|^2$:
\be
\delta_N\left(\delta_N +1\right) - \left|\sigma_a\right|^2 \ge 0.
\label{unc15}
\ee
For the fermion operators, satisfying the relations $\{\hat{b} ,\hat{b}^{\dagger}\}= 1$
and $\hat{b}^2=0$, the following inequality holds:
\be
\delta_N\left(1-\delta_N \right) - |\langle\hat{b}\rangle|^4 \ge 0,
\qquad
\delta_N= \langle \hat{b}^{\dagger}\hat{b}\rangle
- \left|\langle \hat{b}\rangle\right|^2.
\label{unc16}
\ee
Another form of this inequality is
\begin{equation}
\langle\hat b^{\dag}\hat b\rangle
\langle\hat b\hat b^{\dag}\rangle + |\langle\hat b \rangle|^2
\langle\hat b^{\dag}\hat b-\hat b\hat b^{\dag}\rangle
-2|\langle\hat b\rangle|^4\ge 0.
\label{16new}
\end{equation}

A disadvantage of inequality (\ref{unc3}) is that  it often becomes trivial 
for operators different from the coordinate and momentum ones, due
to the very simple reason: for many states and operators the
right-hand side of (\ref{unc3}) equals zero, while the left-hand side is
obviously positive. For example, this is the case for the angular
momentum operators, for which relation (\ref{unc3}) assumes the form
\be
L_{xx} L_{yy} \ge (\hbar^2/4) L_z^2,
\label{unc6}
\ee
where the following notation is used to simplify formulas for operators labeled with indexes:
\[
z_j \equiv \langle \hat{z}_j\rangle,
\quad
z_{jk}=z_{kj}= \frac12 \langle \hat{z}_j\hat{z}_k +
\hat{z}_k\hat{z}_j\rangle
- \langle \hat{z}_j\rangle\langle\hat{z}_k\rangle.
\]
If the average value of operator $\hat{L}_z$ equals zero,
relation (\ref{unc6}) gives no information about the variances $L_{xx}$ and  $L_{yy}$. 
(Nonetheless, inequality (\ref{unc6}) is important, because it tells us that 
 at least one average value $\langle \hat{L}_k\rangle$ with $k \neq j$ 
{\em must\/} be zero for any eigenstate of operator $\hat{L}_j$.)

An insufficient efficiency of relation (\ref{unc6}) may be
explained, partially, by the fact that three equivalent
noncommuting operators $\hat{L}_x$, $\hat{L}_y$ and $\hat{L}_z$
enter this relation on an unequal footing. Therefore, there exists a need
in generalizing inequalities (\ref{unc3}) or (\ref{unc4}) to systems of many (more
than two) operators. This problem was considered for the first time by Robertson \cite{Robertson34}.
His results and some generalizations  are given in the next subsection.

\subsection{Robertson's  inequalities for $N$ arbitrary  operators and their generalizations}
\label{sec-productN}

Considering $N$ arbitrary (not necessarily Hermitian) operators 
$\hat{z}_1$, $\hat{z}_2$,\ldots, $\hat{z}_N$ and following Robertson, 
one can construct the operator
$\hat{f} = \sum_{j=1}^N \alpha_j (\hat{z}_j -\langle \hat{z}_j\rangle)$,
where $\alpha_j$ are arbitrary complex numbers.
All the following results are based on the fundamental 
inequality $\langle\hat{f}^{\dagger}\hat{f}\rangle\ge 0 $, which
must be satisfied for any pure or mixed quantum
state (the symbol $\hat{f}^{\dagger}$ means the Hermitian conjugated operator).
In the explicit form, this inequality is the condition of
positive semi-definiteness of the
quadratic form $\alpha^*_j F_{jm}\alpha_m $
(hereafter the summation over identical indices is assumed),
whose coefficients
$F_{jm} = \langle(\hat{z}_j^{\dagger}- \langle\hat{z}_j\rangle^*)
(\hat{z}_m- \langle\hat{z}_m\rangle)\rangle$
form the Hermitian matrix $F =\Vert F_{jm}\Vert$.
One has only to use the known conditions of
the positive semi-definiteness of Hermitian quadratic forms (see, for
example, \cite{Gantmakher,Bellman,BeckBell,Marshall}) 
to write explicit inequalities containing the elements of 
matrix $F$. All such inequalities can be considered as
generalizations of inequality (\ref{unc3}) to the case of more than two operators. 

If all operators $\hat{z}_j$ are Hermitian, then it is convenient to
split matrix $F$ as $F = X + iY$, where $X$ and $Y$ are
real symmetric and antisymmetric matrices, respectively,
consisting of the elements
\be
X_{mn}= \frac12 \left\langle\left\{\left(\hat{z}_m-
\langle\hat{z}_m\rangle\right),
\left(\hat{z}_n- \langle\hat{z}_n\rangle\right)\right\}\right\rangle,
\label{defX}
\ee
\be
Y_{mn}= \frac1{2i} \left\langle\left[\hat{z}_m,\hat{z}_n\right]
\right\rangle.
\label{defY}
\ee
The symbols $\{,\}$ and $[\,,\,]$ mean, as usual, the anticommutator and the
commutator. 
An important inequality derived by Robertson \cite{Robertson34} reads 
\be
X_{11}X_{22}\ldots X_{NN} \ge  \det X \ge \det Y.
\label{unc10}
\ee
It is worth noting that $\det Y \ge 0$ for any antisymmetric matrix $Y$ 
(although $\det Y=0$ if N is odd number).
Many other inequalities and their geometrical interpretations can be found, e.g., 
in papers \cite{183,Wunsche06,Weig16Math,Qin16}.

For a system of $n$ coordinate and $n$ momentum operators, 
matrix $F$ can be represented as $F = Q - i\hbar\Sigma/2$, where 
$2n \times 2n$ matrices $Q$ and $\Sigma$ consist of  $n \times n$
blocks:
\[
Q = \left\Vert
\begin{array}{cc}
Q_p & Q_{px}\\
Q_{xp} & Q_x
\end{array}
\right\Vert,
\quad
\Sigma = \left\Vert
\begin{array}{cc}
0 & E_n\\
-E_n & 0
\end{array}
\right\Vert,
\]
$E_n$ being the $n \times n$ unit matrix.
Then,
\be
\det( Q_p) \det( Q_x) \ge \left(\hbar^2/4\right)^n.    
\label{unc45}
\ee
The consequence of relation (\ref{unc10}) reads
\be
\label{unc43}  
\left(\overline{x_1^2}\right) 
\cdots \left(\overline{x_n^2}\right)\left(\overline{p_1^2}\right)
\cdots \left(\overline{p_n^2}\right) \ge
\det (Q) \ge \left(\hbar^2/4\right)^n,
\ee
since $\det Y = (\hbar/2)^{2n} \det \Sigma = (\hbar^2/4)^n$.

\subsubsection{Inequalities for the traces of covariance matrices}
\label{traces}

A weakness of the product inequality (\ref{unc3}) is that its left-hand side turns into zero for eigenstates of operators 
$\hat{A}$ ot $\hat{B}$.  Therefore several authors looked for inequalities, whose left-hand sides
contain, instead of products, sums of variances (or their square roots -- ``uncertainties'') of the observables.
One of the first papers in this direction was published by Turner and Snider \cite{Turner81},
who considered the special case of three space dimensions. 
Generalizing  their scheme to $n$ spatial dimensions, 
the following inequality can be obtained \cite{183}:
\be
\mbox{Tr}\left(Q_x\right)\mbox{Tr}\left(Q_p\right)
\ge n^2\frac {\hbar^2}4 +\left[\mbox{Tr}\left(Q_{xp}\right)\right]^2.
\label{61}
\ee
It turns into the equality  for the ground state
of the $n$-dimensional harmonic isotropic oscillator. 

If two observables, $A$ and $B$, have the same physical dimensions, 
an immediate consequence of the Robertson---Schr\"odinger inequality (\ref{unc4}) is 
\be
\sigma_A + \sigma_B \ge 2\left|G_{AB}\right|.
\label{sig+sig}
\ee
Taking the sum of such inequalities with respect to all pairs of observables $z_1, z_2, \ldots,z_N$, 
one can write \cite{Trif02}
\[
\mbox{Tr}(X) \ge  \frac2{N-1} \sum_{1\le j < k \le N} \left|G_{jk}\right| 
\ge \frac2{N-1} \sum_{1\le j < k \le N} \left|Y_{jk}\right|,
\]
with $N(N-1)/2$ terms in the right-hand side.
For the even number of operators $N=2m$, Trifonov \cite{Trif02} obtained the inequality 
containing only $m$ terms in the right-hand side:
\be
\mbox{Tr}(X) \ge  2 \sum_{j=1}^{m}  \left|G_{j,j+m}\right| \ge 2 \sum_{j=1}^{m}  \left|Y_{j,j+m}\right|.
\label{TrX-Gjk-m}
\ee
Another simple inequality was proved in \cite{KW18,unc34,unc5,Chen22}:
\be
\mbox{Tr}(X)  \ge 2\left[\sum_{j<k} Y_{jk}^2\right]^{1/2} 
= \left[2\mbox{Tr}(Y\tilde{Y})\right]^{1/2} .
\label{sumN}
\ee

\subsubsection{Inequalities for sums of uncertainties}
\label{sec-sums}

The following inequality containing the {\em standard deviations\/} (``uncertainties'')
$\Delta A \equiv \sqrt{\sigma_A}$ and $\Delta B \equiv \sqrt{\sigma_B}$ was  found in \cite{Pati07}:
\be
\Delta A + \Delta B \ge \Delta(A+B).
\label{PatiAB}
\ee
Obviously, it is assumed here that observables $A$ and $B$ (described by means of the Hermitian operators)
have the same dimensions, in order that the observable $A+B$ could have a sense. Otherwise, some rescaling
factors should be used. 
A generalization of (\ref{PatiAB}) 
to the case of $N$ observables
and arbitrary real numbers $p_i$ has the form
\be
\sum_{i=1}^N |p_i| \Delta A_i \ge \Delta\left(\sum_{i=1}^N p_i A_i \right).
\label{PatiN}
\ee

Combining (\ref{PatiN}) with the Robertson inequality (\ref{unc3}),
one can arrive at the inequality
\beqn
&&\left(\sum_{j=1}^N |p_j| \Delta A_j \right) \left(\sum_{k=1}^M |q_k| \Delta B_k \right) 
\nonumber \\ &&
\ge 
\frac12 \left|\sum_{j=1}^N \sum_{k=1}^M p_j q_k \left\langle \left[\hat{A}_j, 
\hat{B}_k\right] \right\rangle \right|.
\label{PatiABC}
\eeqn
For $N$ coordinates and momenta one obtains \cite{Pati07}
\be
\left(\sum_{i=1}^N  \Delta x_i \right) \left(\sum_{i=1}^N  \Delta p_i \right) \ge N\hbar/2.
\label{Pati-xp}
\ee

\subsubsection{Symplectic invariants and uncertainty relations}

In view of complicated explicit structures of inequalities expressing multidimensional uncertainty relations,
 several authors  \cite{Simon94, Sudar95} studied possible {\em canonical forms\/} 
of the $2n \times 2n$
covariance matrices. If the initial momentum-coordinate vector
${\bf q} =\left( p_1, p_2, \ldots, p_n, x_1, x_2, \ldots, x_n\right)$ 
is linearly transformed as  ${\bf q}=S{\bf q}^{\prime}$,
then, the covariance matrix $Q$ is related to the transformed matrix  $Q^{\prime}$ as $Q= SQ^{\prime}\tilde{S}$,
where $\tilde{S}$ is the transposed matrix. 
Since the uncertainty inequalities are determined by the commutator matrix $Y$,
it seems reasonable to use transformation that does not change matrix $Y= -i(\hbar/2)\Sigma$. Such transformations,
satisfying the condition $S\Sigma \tilde{S} =\Sigma$, are called {\em symplectic transformations}. 
In particular, $|\det S| =1$, so that $\det Q = \det Q^{\prime}$.
The fundamental theorem
in this area, proved by Williamson \cite{Williamson} (see also \cite{Arnold,Simon99} for the discussion and simplified proofs), 
tells us that any positive definite symmetrical matrix $Q$ 
can be transformed by means of symplectic transformations to the canonical diagonal form
$
Q^{(can)}=\mbox{diag}\left(\kappa_1, \kappa_2, \ldots, \kappa_n, \kappa_1, \kappa_2, \ldots, \kappa_n\right)$,
with positive values $\kappa_j$. The uncertainty relation in this formulation is the statement that 
\be
\kappa_j \ge \hbar/2, \qquad \mbox{for all} \; j=1,2,\ldots,n.
\label{unc-kapj}
\ee
 (A reduction to the diagonal form with identical blocks $Q_x$ and $Q_p$ implies some scaling
transformations to arrive at blocks with the same physical dimension.)
Returning to the initial $2n\times 2n$ matrix $Q$ with $n\times n$ block, one can
obtain  \cite{Sudar95,Trif97} the following consequences of (\ref{unc-kapj}):
\be
\mbox{Tr}\left[(iQ\Sigma)^{2k}\right] \ge 2^{1-2k} n\hbar^{2k}, \qquad k=1,2,\ldots.
\label{Sud-tr}
\ee
\be
\mbox{Tr}\left(Q_p Q_x\right)  - \mbox{Tr}\left(Q_{px}^2\right) \ge n\hbar^2/4.
\label{TrQpQx}
\ee

Important inequalities can be obtained if one considers the following polynomial of order $2n$
with respect to an auxiliary parameter $\mu$:
\[
{\cal D} (\mu) \equiv \det(Q -\mu\Sigma) = \sum_{k=0}^{2n} {\cal D}^{(n)}_k \mu^k .
\]
This polynomial is invariant with respect to any symplectic transformation. Consequently, each coefficient
${\cal D}^{(n)}_k$ is invariant with respect to such transformations, as well. Therefore, the coefficients
${\cal D}^{(n)}_k$ were named in \cite{183,univ00} ``{\em quantum universal invariants}'',
because their values are preserved in time during the evolution governed by {\em arbitrary quadratic Hamiltonians}.

After the reduction of matrix $Q$ to the canonical diagonal form, one can write
$
{\cal D}(\mu) = \prod_{j=1}^n \left(\kappa_j^2 +\mu^2\right)$.
Consequently, ${\cal D}(\mu)={\cal{D}}(-\mu)$, and the only nonzero universal invariants ${\cal D}^{(n)}_{2j}$
can be expressed in terms of the symplectic eigenvalues $\kappa_j$ as follows:
\[
{\cal D}^{(n)}_{0} = \left(\kappa_1 \kappa_2\ldots\kappa_n\right)^2,  
\quad
{\cal D}^{(n)}_{2} = {\cal D}^{(n)}_{0}\sum_{j=1}^n \kappa_j^{-2}, 
\]
\[
{\cal D}^{(n)}_{4} = {\cal D}^{(n)}_{0}\sum_{j<k}\left( \kappa_j\kappa_k\right)^{-2},  \ldots
\]
\[
{\cal D}^{(n)}_{2n-4} = \sum_{j<k}\left( \kappa_j\kappa_k\right)^{2}, \quad
{\cal D}^{(n)}_{2n-2} = \sum_{j=l}^n  \kappa_j^{2}. 
\]
Obviously, minimal values of all these expressions can be achieved for 
$\kappa_1=\kappa_2=\cdots=\kappa_n=\hbar/2$,
resulting in the following set of inequalities \cite{Sudar95}:
\be
{\cal D}^{(n)}_{2j} \ge \left(\hbar^2/4\right)^{n-j}\frac{n!}{j!(n-j)!},
\label{D2jgen}
\ee
\be
{\cal D}^{(n)}_{2n-2} \equiv \sum_{i,j=1}^n\left(\overline{p_ip_j}
\cdot\overline {x_ix_j}-\overline {p_ix_j}\cdot\overline{x_ip_j}\right) 
\ge \frac{\hbar^2 n}{4}. 
\label{D2j>big}
\ee
The following notation is used here:
\be
\overline{A} \equiv \langle\hat{A}\rangle, \qquad
\overline{BC} \equiv \frac12\langle \hat{B} \hat{C} + \hat{C}\hat{B}\rangle
- \overline{B}\cdot\overline{C}.
\label{unc32}
\ee

\subsection{Inequalities for three operators}
\label{sec-3op-gen}

The case of three Hermitian operators $\hat{z}_j$ ($j=1,2,3$) was studied in detail for
the first time by Synge \cite{Synge71}. 
One of his main results is the inequality
\beqn
3 X_{11}X_{22}X_{33} &>&  X_{11}\left(X_{23}^2 + Y_{23}^2\right)
+X_{22}\left(X_{13}^2 + Y_{13}^2\right) 
\nonumber \\ &&
+X_{33}\left(X_{12}^2 + Y_{12}^2\right),
\label{unc18}
\eeqn
where the equality cannot be reached for any quantum state.

Applying the known inequality for the
arithmetic mean and the geometric mean \cite{Marshall} 
\be
\sum_{k=1}^n x_k \ge n\left(\prod_{k=1}^n x_k\right)^{1/n}, \qquad x_k >0
\label{arif-geom}
\ee
with $n=3$ to the right-hand side of (\ref{unc18}), one can arrive at the inequality 
\be
 X_{11}^2 X_{22}^2 X_{33}^2 \ge  \left(X_{23}^2 + Y_{23}^2\right)
\left(X_{13}^2 + Y_{13}^2\right) \left(X_{12}^2 + Y_{12}^2\right).
\label{unc18-new}
\ee

\subsubsection{Inequalities without covariances}
\label{subsec-Clifford}

An obvious disadvantage of inequalities like (\ref{unc10}) is that they are 
rather complicated for $N>2$ observables,
 because they contain, in addition to $N$ variances $X_{kk}$ 
and $N(N-1)/2$ mean values of commutators $Y_{jk}$,
numerous sums and products of various combinations of $N(N-1)/2$ covariances $X_{jk}$ with $j\neq k$. 
 For example, if $N=4$, then $\det X$ contains $17$ different products of covariances 
(the explicit expression can be found in Ref. \cite{univ00})
in addition to $6$ different products of mean values of commutators in $\det Y$.
Moreover, inequality (\ref{unc10}) seems totally useless if $N$ is an odd number, as soon as
$\det Y=0$ in this case.

One can get rid of all $N(N-1)/2$ covariances, using the scheme proposed in paper \cite{unc5}. 
Suppose that we know $N$ Hermitian $M\times M$ anticommuting matrices $R_k$
satisfying the relations of the Clifford algebra
\be
R_j R_k +R_k R_j = 2 I_M \delta_{jk},
\label{Rjk}
\ee
where $I_M$ is the $M\times M$ unit matrix.
Consider the operator $\hat{f} = \sum_{k=1}^N \xi_k \delta\hat{z}_k R_k$ with 
 arbitrary {\em real\/} coefficients $\xi_k$ and arbitrary Hermitian operators $ \hat{z}_k$.
It acts in the extended Hilbert space of states
$|\Psi\rangle =|\psi\rangle\otimes |\chi\rangle$, where $|\chi\rangle$ is an auxiliary $M$-dimensional vector.
Then,  the condition $\langle\Psi| \hat{f}^{\dagger}\hat{f}|\Psi\rangle \ge 0$ 
can be written as the condition of
positive semi-definiteness of the Hermitian $M\times M$ matrix
$
F = g I_M + i\sum_{j<k}  R_j R_k y_{jk}$,
where
$
 g = \sum_{k=1}^N \xi_k^2 X_{kk}$ and $y_{jk} = 2\xi_j \xi_k Y_{jk} $.
The covariances $X_{jk}$ with $j \neq k$ go out due to the anti-commutation relations
(\ref{Rjk}).

To perform the scheme, one has to know the explicit form of $N$ anticommuting matrices $R_j$, satisfying the
Clifford algebra relations (\ref{Rjk}). The main technical problem is the dimension of such matrices:
$2^n \times 2^n$ for $N=2n$ and $N=2n+1$ \cite{Okubo}.
Therefore, I confine myself here to
the special case of $N=3$, when matrices $R_j$ are three Pauli's $2\times2$ matrices $\sigma_j$
\cite{unc34}. 
The cases of $N=4$ and $N=5$ (four Dirac's $4\times4$ matrices) were studied in papers
\cite{unc34,unc5}.
Using the properties of the Pauli matrices and performing averaging over the state $|\psi\rangle$, 
one can write
$
\langle\Psi| \hat{f}^{\dagger}\hat{f}|\Psi\rangle =
\langle\chi\ | {\cal A} |\chi\rangle
$
with the $2\times2$ Hermitian matrix which
 does not contain the covariances $X_{jk}$ with $j \neq k$
(here $\sigma_0$ is the $2\times2$ unit matrix):
\beqn
{\cal A} &=& \left(\alpha_1^2 X_{11} + \alpha_2^2 X_{22} +\alpha_3^2 X_{33}\right)\sigma_0
\nonumber \\ 
&-& 2 \sigma_1 \alpha_2 \alpha_3 Y_{23} -2 \sigma_2 \alpha_3 \alpha_1 Y_{31} -2 \sigma_3 \alpha_1 \alpha_2 Y_{12}.
\label{calA}
\eeqn
The condition $\det{\cal A} \ge 0$, which guarantees that  matrix ${\cal A}$ is
 positive semi-definite, yields the inequality  \cite{unc34}
\beqn
&&\alpha_1^2 X_{11} +\alpha_2^2 X_{22} + \alpha_3^2 X_{33} 
\nonumber \\ &&
\ge
2\left[\left(\alpha_1 \alpha_2 Y_{12}\right)^2 +\left(\alpha_2 \alpha_3 Y_{23}\right)^2 
+\left(\alpha_1 \alpha_3 Y_{13}\right)^2
\right]^{1/2},
\label{gen}
\eeqn
which must hold for {\em arbitrary real numbers\/} $\alpha_1$, $\alpha_2$ and $\alpha_3$.
The choice $\alpha_1=\alpha_2=\alpha_3$ results in the inequality
\be
 X_{11} + X_{22} +  X_{33} \ge
2\left[ Y_{12}^2 + Y_{23}^2 + Y_{13}^2
\right]^{1/2}.
\label{sum3}
\ee
Choosing $\alpha_k^2 =X_{kk}^n$, one obtains the inequality
\beqnn
 && X_{11}^{n+1} + X_{22}^{n+1} +  X_{33}^{n+1} 
\\ &&
\ge
2\left[ Y_{12}^2 X_{11}^n X_{22}^n + Y_{23}^2X_{33}^n X_{22}^n + Y_{13}^2X_{11}^n X_{33}^n
\right]^{1/2}.
\eeqnn
Wishing to find an inequality for the triple product $X_{11}  X_{22}   X_{33}$, let us choose
$\alpha_1^2 =X_{22}X_{33}$, $\alpha_2^2 =X_{11}X_{33}$ and $\alpha_3^2 =X_{22}X_{11}$.
Then, the following inequality arises:
\be
X_{11}  X_{22}   X_{33} \ge \frac49 \left(X_{11}Y_{23}^2  + X_{22}Y_{13}^2 +X_{33}Y_{12}^2\right).
\label{prod3}
\ee
Applying  inequality (\ref{gen}) with $\alpha_1=Y_{23}$,
$\alpha_2=Y_{31}$ and $\alpha_3=Y_{12}$ to the right-hand side of (\ref{prod3}), 
one can obtain the inequality  \cite{unc5}
\be
X_{11}X_{22}X_{33} \ge (4/3)^{3/2}\left|Y_{12}Y_{13}Y_{23}\right|.
\label{3prod}
\ee

\subsubsection{Special cases}
\label{sec-special}

A natural system
of three operators is the set of the angular momentum 
operators $\hat{L}_x, \hat{L}_y, \hat{L}_z$. 
A lot of inequalities for these operators have been found in Refs. 
\cite{183,Delbourgo77,Rivas08, Damm15,Bjork16}. I bring here few simple examples.

The generalization of inequality (\ref{unc6}) reads
\be
L_{xx}L_{yy} - L_{xy}^2 \ge \frac{\hbar^2}{4}L_z^2.
\label{unc29}
\ee
Inequality (\ref{sum3}) can be written as
\be
\langle {\bf L}^2\rangle \ge \langle {\bf L}\rangle^2 + \hbar\left|\langle {\bf L}\rangle\right|,
\label{L2L}
\ee
where the second term in the right-hand side appears due to the non-commutativity
of components of the operator vector ${\bf {L}}$.
A consequence of (\ref{unc6}) is the inequality
\be
L_{xx}L_{yy} +L_{yy}L_{zz} +L_{zz}L_{xx} \ge \frac{\hbar^2}{4}\langle \hat{\bf L}\rangle^2.
\label{Delta2}
\ee
Moreover, if 
$L_{zz} \ge L_{xx} \ge L_{yy}$, then
\be
L_{zz}\left(L_{xx} + L_{yy}\right) \ge \frac{\hbar^2}4 \left\langle \hat{\bf L}\right\rangle ^2.
\ee

In the case of spin-$1/2$ operators,
satisfying the anticommutation relation
$\{\hat{s}_{\alpha}\,,\, \hat{s}_{\beta}\} = \delta_{\alpha\beta}\hbar^2/2 $,
the (co)variances $s_{jk}$ depend on the mean values $s_j$ and $s_k$ only:
$s_{jk}=  \delta_{jk}\hbar^2/4 -s_j s_k$, where $s_{j} \equiv \langle\hat{s}_{j}\rangle$.
 Since $s_{jj} \ge 0$, the simplest inequality is
$s_j^2 \le \hbar^2/4$. The sum of three such inequalities yields $s_x^2 + s_y^2 + s_z^2 \le 3\hbar^2/4$.
However, the correct inequality, following from (\ref{unc29}), is much stronger:
\be
s_x^2 + s_y^2 + s_z^2 \le \frac{\hbar^2}{4}.
\label{unc30}
\ee

Returning to the coordinate and momentum operators in one
dimension, one may introduce three operators,
\[
\hat{R}_1 = \left(\delta\hat{p}\right)^2, \quad
\hat{R}_2 = \left(\delta\hat{x}\right)^2, \quad
\hat{R}_3 = \frac12\left(\delta\hat{p}\delta\hat{x}
+\delta\hat{x}\delta\hat{p}\right),
\]
where
$\delta\hat{p} = \hat{p} -\langle\hat{p}\rangle$ and
$\delta\hat{x} = \hat{x} -\langle\hat{x}\rangle$.
The following inequalities were derived (together with many others) in Ref. \cite{183}:
\[
R_{11}R_{22} - R_{12}^2 \ge 4\hbar^2 R_3^2, \qquad
R_1 R_2 - R_3^2 \ge \hbar^2/4,
\]
\[
R_{11}R_{33} - R_{13}^2 \ge \hbar^2 R_1^2, \qquad
R_{22}R_{33} - R_{23}^2 \ge \hbar^2 R_2^2.
\]
\beqnn
&&\left[\overline {(\delta p)^4}-\sigma_{pp}^2\right]
\left[\overline {(\delta x)^4}-\sigma_{xx}^2\right]
\ge
\hbar^4 +6\hbar^2\sigma_{px}^2
\\ &&
- 2\overline {(\delta p\delta x)^2} \left(\sigma_{pp}\sigma_{xx} +\frac34\hbar^2\right),
\eeqnn
where 
$
\overline {(\delta p\delta x)^2}= \frac14\left\langle\left[\left
(\delta\hat p\delta\hat x+\delta\hat x\delta\hat p
\right)\right]^2\right\rangle$.

\subsection{Bargmann--Faris inequalities}

Interesting multidimensional extensions of inequality (\ref{unc2}) were found by Bargmann \cite{Barg72}
and generalized by Faris \cite{Faris}.
 The starting point is the inequality
for arbitrary operators $\hat{A}$ and $\hat{B}$
\be
\langle\hat{A}^{\dagger}\hat{A}\rangle + \langle\hat{B}^{\dagger}\hat{B}\rangle
\ge i \langle\hat{A}^{\dagger}\hat{B} - \hat{B}^{\dagger}\hat{A}\rangle,
\label{149}
\ee
 which follows from the inequality 
$\langle (\hat{A} - i\hat{B})^{\dagger}(\hat{A} - i\hat{B})\rangle \ge 0$. 
 Replacing $\hat{A}$ by $\hat{A}\langle\hat{A}^{\dagger}\hat{A}\rangle^{-1/2}$
 and $\hat{B}$ by $\hat{B}\langle\hat{B}^{\dagger}\hat{B}\rangle^{-1/2}$, 
 one obtains a special case of inequality (\ref{unc14}):
\be
2\left[\langle\hat{A}^{\dagger}\hat{A}\rangle  \langle\hat{B}^{\dagger}\hat{B}\rangle
\right]^{1/2}
\ge i \langle\hat{A}^{\dagger}\hat{B} - \hat{B}^{\dagger}\hat{A}\rangle.
\label{150}
\ee
Another useful inequality follows from the Schwarz inequality
\be
\left[\langle\hat{A}^{\dagger}\hat{A}\rangle  \langle\hat{B}^{\dagger}\hat{B}\rangle
\right]^{1/2}
\ge \left| \langle\hat{A}^{\dagger}\hat{B} \rangle\right|,
\label{151}
\ee
if one assumes that $\hat{A} = \hat{Y}^{1/2}$ and $\hat{B} = \hat{Y}^{-1/2}$,
where $\hat{Y}$ is a positively definite operator:
\be
\langle\hat{Y}\rangle \ge  \langle\hat{Y}^{-1}\rangle^{-1}.
\label{152}
\ee

 To obtain a multidimensional generalization of inequality (\ref{unc2}), 
let us separate the angular and radial parts of the operator 
$\hat{\bf p}^2 = -\hbar^2\Delta$,
 introducing the angular momentum projection operators 
 $\hat{L}_{jk}=\hat{x}_j\hat{p}_k -\hat{x}_k\hat{p}_j$
 in the $n$-dimensional case. Each operator $\hat{L}_{jk}$ commutes with the operators 
 $\hat{\bf x}^2 = \hat{x}_1^2 + \hat{x}_2^2 + \ldots + \hat{x}_n^2 
\equiv \hat{r}^2$
and $\hat{\bf p}^2$, and the same is true for the operator 
$\hat{\bf L}^2 =\sum_{j<k} \hat{L}_{jk}^2$.
 One can check the relations
\be
\hat{r}^{-2}\hat{\bf L}^2 = \frac12 \sum_{j,k} \hat{L}_{jk}\hat{r}^{-2} \hat{L}_{jk}
= \hat{\bf p}^2 -\left(\hat{\bf p}\hat{\bf x}\right)\hat{r}^{-2}
\left(\hat{\bf x}\hat{\bf p}\right),
\label{153}
\ee
which can be written in two equivalent forms:
\be
 \hat{\bf p}^2 = \hat{r}^{-2}\hat{\bf L}^2 +\hat{A}_1^{\dagger}\hat{A}_1
 = \hat{r}^{-2}\hat{\bf L}^2 +\hat{A}_2^{\dagger}\hat{A}_2,
\label{154}
\ee
where
 $\hat{A}_1 = \hat{r}^{-1}\left(\hat{\bf x}\hat{\bf p}\right)$
 and $\hat{A}_2 = \hat{r}\left(\hat{\bf p}\hat{\bf x}\right)\hat{r}^{-2}$.
Then, using inequality (\ref{150}) with
  $\hat{A} = \hat{A}_1$ or $\hat{A} = \hat{A}_2$ and $\hat{B}=\varphi(r)$,
  where $\varphi(r)$ is
an arbitrary real sufficiently smooth function without singularities
(except at the point $r = 0$), one can arrive at the inequality
\be
\hbar\left\langle \frac{\partial\varphi}{\partial r} +\nu_k \frac{\varphi(r)}{r}\right\rangle
\le 2\left[\langle \varphi^2(r)\rangle
\langle \hat{\bf p}^2 - \hat{r}^{-2}\hat{\bf L}^2\rangle\right]^{1/2},
\label{156}
\ee
where the value $\nu_1=n-1$ corresponds to the choice $\hat{A} = \hat{A}_1$
and the value $\nu_2=3-n$ corresponds to $\hat{A} = \hat{A}_2$.

Assuming $\varphi(r) = r^{1+\alpha}$, one obtains the inequality \cite{Faris}
\be
\hbar(1+\alpha+\nu_k)\langle r^{\alpha}\rangle
\le 2\left[\langle r^{2+2\alpha}\rangle
\langle \hat{\bf p}^2 - \hat{r}^{-2}\hat{\bf L}^2\rangle\right]^{1/2}.
\label{158}
\ee
Taking $\alpha = 0$ and choosing $\nu=\nu_1$, one arrives at
 the special case of inequality (\ref{61}):
\be
\left[\langle r^{2}\rangle
\langle \hat{\bf p}^2 \rangle\right]^{1/2}
\ge n\hbar/2.
\label{159}
\ee
If $\alpha = -2$, then, $1 + \alpha + \nu_k = \pm(n - 2)$, 
and the consequences of (\ref{158}) are the inequalities
\be
\langle\hat{\bf p}^2 \rangle \ge 
\langle \hat{\bf p}^2 - \hat{r}^{-2}\hat{\bf L}^2\rangle
\ge \frac{\hbar^2}{4}(n-2)^2\langle r^{-2}\rangle.
\label{161}
\ee
If $\alpha = -1$, Eq. (\ref{158}) leads to the inequalities (for $n \ge 2$)
\be
\langle\hat{\bf p}^2 \rangle \ge 
\langle \hat{\bf p}^2 - \hat{r}^{-2}\hat{\bf L}^2\rangle
\ge \left[\frac{\hbar}{2}(n-1)\langle r^{-1}\rangle\right]^2.
\label{162}
\ee

The following inequality was proven
for arbitrary real functions of several variables in study \cite{Baum}:
\be
\langle\hat{\bf p}^2 \rangle \langle \varphi^2({\bf x})\rangle 
\ge \left({\hbar^2}/{4}\right)\langle \nabla \varphi({\bf x})\rangle^2.
\label{168}
\ee
 A generalization of relation
(\ref{168}) 
was proven in study \cite{Ex84}:
\[
\langle\hat{\bf p}^2 \rangle \langle \varphi^2({\bf x})\rangle 
\ge ({\hbar^2}/{4})\left\langle r^{-1}\left[{\bf x}\nabla \varphi({\bf x})
+(n-1) \varphi({\bf x})\right] \right\rangle^2.
\]
Besides, the following inequality holds for quantum states
 with a given value of the angular momentum $l$  \cite{Ex84}:
\[
\langle\hat{\bf p}^2 \rangle \langle \varphi^2(r)\rangle 
\ge ({\hbar^2}/{4})\left\langle {\partial\varphi}/{\partial r}
+(n-1 +2l) \varphi(r)/r \right\rangle^2.
\]
Its special case is the inequality \cite{Barg72}
\be
\langle\hat{\bf p}^2 \rangle \langle r^{2+2\alpha}\rangle 
\ge ({\hbar^2}/{4})\left[(n+\alpha+2l)\langle r^{\alpha}\rangle \right]^2.
\label{167}
\ee
The case of $\alpha=0$ was also considered in Ref. \cite{Moreno06}:
\be
\langle\hat{\bf p}^2 \rangle \langle r^{2}\rangle 
\ge ({\hbar^2}/{4})(n+2l)^2.
\ee
Taking $\alpha=-2$, one obtains the following analog of (\ref{161}):
\be
\langle\hat{\bf p}^2 \rangle 
\ge \frac{\hbar^2}{4}(n-2 +2l)^2\langle r^{-2}\rangle.
\label{161-l}
\ee

The following inequalities were derived in \cite{Bracher11} for the states with  fixed
angular momentum quantum numbers in two and three dimensions:
\be
\Delta r \Delta p \ge \left\{
\begin{array}{ll}
\hbar(|m|+1) & \mbox{in}\,\, 2D
\\
\hbar(l+3/2) & \mbox{in}\,\,  3D
\end{array}
\right. .
\label{Brac}
\ee
Note that all relations of this section remain valid if one makes the replacement of
vectors ${\bf r} \leftrightarrow {\bf p}$.

\subsubsection{Applications to the hydrogen atom}
\label{sec-BF-H}

The ground state energy of the $n$-dimensional ``hydrogen atom'' can be found with the aid
of inequality (\ref{162}): 
\be
E \ge \frac{\hbar^2}{2m}\lambda_n^2 \xi^2 - e^2 \xi, \quad \xi =\langle r^{-1}\rangle,
\quad \lambda_n = \frac12(n-1).
\label{(164)}
\ee
Minimizing the right-hand side of (\ref{(164)}) with respect to parameter $\xi$, one obtains the
{\em exact\/} minimal energy 
\be
E_{min} = -me^4/(2\hbar^2\lambda_n^2).
\ee
 The relations (\ref{149})-(\ref{150})
turn into equalities for states satisfying the condition 
\be
\left(\hat{A} -i\hat{B}\right) |\psi\rangle =0 .
\label{(165)}
\ee
In the case of inequality (\ref{162}), we have $\hat{A} = \hat{A}_1 = -i\hbar\partial/\partial r$ and  
$\hat{B} = \beta = const$. Therefore, the solution to Eq. (\ref{(165)}) is the function 
$\psi(r)= C \exp(-\beta r/\hbar)$.
Choosing the parameter $\beta$ in order to satisfy the condition 
$\langle r^{-1}\rangle = \xi_{min} = me^2/(\hbar^2\lambda_n^2)$
[corresponding to the minimum of the right-hand side of (\ref{(164)})],
one obtains not only the
energy of the hydrogen atom ground state, but also its wave function
\cite{Barg72,Faris}. 

For excited states with $\langle\hat{L}^2\rangle =\hbar^2 l(l+1)$, 
 it is
possible to obtain exact values of energies, using  the
inequality (\ref{167}) with $\alpha = -1$.
Then, parameter $\lambda_n$ in Eq. (\ref{(164)}) should be replaced by $\lambda_n +l$, resulting in
the exact lowest energy for the given value of the angular momentum  $E_{n,l} = -\mbox{Ry}/(\lambda_n +l)^2$.
Remember that $n$ is the space dimension here.

 It was shown in \cite{Barg72} that the equality in (\ref{167}) with $\alpha= -1$ is attained 
for functions of the form
\[
\psi_l({\bf x}) = r^l Y_l ({\bf x}/r) \exp\left(-r/r_l \right), \quad 
r_l = \left(\lambda_n +l\right)\frac{\hbar^2}{me^2}
\]
 (where $Y_l ({\bf x}/r)$ is a spherical harmonic),
i.e., the hydrogen eigenfunctions describing Bohr's orbits.

\subsection{Inequalities for higher moments}

A special case of the standard uncertainty relation (\ref{unc2}) is the inequality
\be
\langle \hat{x}^2\rangle \langle \hat{k}^2\rangle \ge \frac14 = {\cal P}_2, \quad
\hat{k}= \hat{p}/\hbar.
\label{Pi2}
\ee
Introducing the notation $\Pi^{(n)}\equiv\langle |\hat{x}|^n\rangle \langle |\hat{k}|^n\rangle$,
it seems natural to look for the minimal value of this product, 
$
{\cal P}_n = \mbox{min}\left\{\Pi^{(n)}\right\}$,
for an arbitrary fixed exponent $n$.
A trivial estimation can be made with the aid of (\ref{Pi2}) and the inequality
$\langle \hat{A}^2\rangle \ge \langle \hat{A}\rangle^2$:
${\cal P}_4 \ge 4^{-2}$.
However, this estimation is very weak.
Indeed, the left-hand side of the
 relation (\ref{unc2}) is minimal for the Gaussian
states. But the fourth-order moments in these states  are given by the formula
$\langle \hat{x}^4\rangle = 3\langle \hat{x}^2\rangle^2$. Hence,
 the product of the fourth-order moments in the 
Gaussian states equals 
$\langle \hat{k}^4\rangle \langle \hat{x}^4\rangle = 9/16$.
This value is nine times higher than the trivial lower bound. 

Several authors looked for the minimal value of the product $\Pi^{(4)}$ using various numeric schemes.
The best known minimal value $\Pi^{(4)}_{min}=0.4878$ was found in paper
\cite{Lynch90} for some finite superposition of the Fock states
of the form $\sum_{n=0}^{K}c_n|4n\rangle$ with $K=6$.
Practically the same minimal value $\Pi^{(4)}_{min} \approx 0.49$ was found
in paper \cite{CDD20} for the superposition of four
coherent states with equal amplitudes and phases shifted by $\pi/2$,
\beqnn
|\psi\rangle_{4\alpha} &=& B\left(|\alpha\rangle  +|i\alpha\rangle +|-\alpha\rangle +|-i\alpha\rangle \right)
\\
&=& {\cal N} \sum_{n=0}^{\infty} \frac{\alpha^{4n}}{\sqrt{(4n)!}} |4n \rangle,
\eeqnn
with $\alpha\approx 0.67$. 
No good analytical bounds for the products ${\cal P}_n$ with $n >2$ have been found until now.

\section{Entropic Uncertainty Relations}
\label{sec-entropic}

The mathematical significance of inequality (\ref{unc2}) is that the
distribution functions of coordinates $|\psi(x)|^2$ and momenta $|\varphi(p)|^2$
cannot be simultaneously localized in arbitrary small domains, if
$\psi(x)$ and $\varphi(p)$ are related by means of the Fourier transform,
\be
\varphi(p)=(2\pi\hbar)^{-1/2}\int \psi(x)\exp(-ipx/\hbar)\,dx.
\label{unc103}
\ee
It
appears that this statement can be also expressed mathematically with the
aid of other inequalities, which contain, instead of variances,
 the quantities ($k \equiv p/\hbar$)
\be
S_x = -\int |\psi(x)|^2\ln(|\psi(x)|^2)\,dx,
\label{unc104x}
\ee
\be
S_k = -\int |\varphi(k)|^2\ln(|\varphi(k)|^2)\,dk.
\label{unc104k}
\ee
 They can be called ``the coordinate distribution entropy'' and
``the momentum distribution entropy'', respectively. 
One should remember that neither ``entropies'' $S_x$
and $S_k$ have anything in common (except the name) with the true
quantum mechanical entropy, defined according to formula 
$
S= -\mbox{Tr}(\hat\rho\ln \hat\rho)$.
 Moreover, the true entropy exactly equals zero for
pure quantum states considered in the present section, while the
integrals (\ref{unc104x}) and (\ref{unc104k}) are different from zero, as a rule. 

Suppose the
function $|\psi(x)|^2$ to be localized in some small domain. Then, the
values of $|\psi(x)|^2$ inside this domain are large, so that
$\ln(|\psi(x)|^2) >0$.
 As a result, $S_x < 0$, since the contribution
to the integral of the points outside the localization domain are
suppressed by small values of $|\psi(x)|^2$ in those points. If the
function $|\varphi(k)|^2$ were also highly localized,
then the relation $S_k < 0$ would be fulfilled, too, resulting in the inequality $S_x + S_k < 0$.
However, this is impossible, due to the inequality proven for the
first time by Hirschman \cite{Hirschman} and Bourret \cite{Bourret}:
\be
S_x + S_k \ge \ln(2\pi).
\label{105}
\ee
Note that definitions (\ref{unc104x}) and  (\ref{unc104k}) 
contain some ambiguities, because functions
$\psi(x)$ and $\varphi(k)$ are not dimensionless. To give sense to
the term $\ln|\psi|^2$, one has to introduce some length scale $x_0$ and
imply the dimensionless variable $x/x_0$ whenever the coordinate $x$
appears. The value of $S_x$ in such a case depends on the choice of the
parameter $x_0$. However, since the corresponding scale in the wave
number space equals $k_0 = x_0^{-1}$, the actual value of $x_0$ in inequality
(\ref{105}) cancels out, so that no ambiguity arises {\em for the sum of two entropies}.

The advantage of relation (\ref{105}) becomes quite clear, if one considers
an example of a ``two-hump'' function $|\psi(x)|^2$, 
represented by two narrow peaks, whose centers are separated by a large distance,
e.g., 
\[
\psi(x)= N\left\{\exp\left[-(x-a)^2/b^2\right] + \exp\left[-(x+a)^2/b^2\right]\right\}
\]
with $ |b| \ll |a|$.
 Although both peaks can be very narrow if $|b| \to 0$, the variance $\sigma_x$ can be
made as large as desired, simply by increasing the distance $2|a|$ between the peaks. The obvious
disadvantage of inequality (\ref{unc2}) is that it does not forbid a
possibility of the existence of a narrow hump of the function $|\varphi(k)|^2$,
when the small variance of $\sigma_p$ would be compensated for by a large
value of the variance $\sigma_x$. In fact, this is impossible, 
and inequality (\ref{105}) shows it quite distinctly: ``the coordinate entropy'' $S_x$
is practically unchanged when the humps are moved apart, i.e., it
remains negative. Therefore, the value of $S_k$ obviously must be
positive, so that the function $|\varphi(k)|^2$ cannot assume large values
anywhere.

For any uncorrelated Gaussian state (satisfying the additional restriction $\sigma_{xp}=0$),
the following relations hold:
\[
S_x =\frac12\ln(2\pi e \sigma_x), \quad S_k =\frac12\ln(2\pi e \sigma_k). 
\]
Since $\sigma_x\sigma_k=1/4$ for such states, $S_x + S_k = \ln(\pi e)$.
Therefore, as long ago as in Hirschman's paper \cite{Hirschman}, a conjecture was
made that, in fact, the right-hand side of (\ref{105}) should be replaced by the
quantity $\ln(\pi e)$. It was proved at the ``physical level of rigor''
by Leipnik \cite{Leipnik59}, while 
strict mathematical proofs were given only fifteen years later in studies \cite{Beckner,Bir-Myc},
using rather heavy calculations: 
\be
S_x +S_k \ge  n \ln (\pi e).
\label{109}
\ee
Here, $n$ is the dimension of the coordinate space.

It is essential that inequality (\ref{109}) is
{\em stronger\/} than the Heisenberg inequality (\ref{unc2}), in the sense that
relation (\ref{unc2}) is a {\em consequence\/} of (\ref{109}). Indeed, looking for the
extremum of the functional $S_x(\{\psi\})$ with a given norm $\Vert\psi\Vert_2$ and a
given variance $\sigma_x =\left\langle\left(\hat{\bf x} -\langle \hat{\bf x}\rangle\right)^2\right\rangle$,
one can easily find that the
``entropy'' $S_x$ is maximal for the Gaussian states of the form 
\[
|\psi({\bf x})|^2 = (2\pi \sigma_x/n)^{-n/2} \exp\left[ -n\left({\bf x} -\langle \hat{\bf x}\rangle\right)^2
/2\sigma_x \right],
\]
so that
\[
S_x \le \frac{n}{2}\ln(2\pi e\sigma_x/n), \quad
S_k \le \frac{n}{2}\ln(2\pi e\sigma_k/n).
\]
Combining these inequalities with (\ref{109}), one obtains
\be
\frac{2\sigma_k}{n} \ge \frac1{\pi e} \exp\left(\frac{2}{n}S_k\right) \ge
(\pi e) \exp\left(-\,\frac{2}{n}S_x\right) \ge \frac{n}{2\sigma_x}.
\label{112}
\ee
Inequality (\ref{unc2}) follows from (\ref{112}) for $n= 1$. The consequence of
(\ref{112}) for $n\ge 2$ is the special case of inequality (\ref{61}) with $\mbox{Tr}(Q_{xp}) = 0$,
since $\sigma_x \equiv \overline{x_1^2} + \cdots
+ \overline{x_n^2} \equiv \mbox{Tr}(Q_x)$.

 The following ``entropic'' inequality,
relating the angular momentum component $L_z$ and the polar angle $\varphi$ in the plane $xy$, was
given in study \cite{Bir-Myc}:
\be
-\sum_{m=-\infty}^{\infty} |c_m|^2 \ln(|c_m|^2) -\int_0^{2\pi} \frac{d\varphi}{2\pi}
|F(\varphi)|^2 \ln(|F(\varphi)|^2) \ge 0.
\label{113}
\ee
Here, $c_m$ are complex Fourier coefficients of the wave function $F(\varphi)$ in the polar coordinates:
$
F(\varphi) = \sum_{m=-\infty}^{\infty} c_m e^{im\varphi}$.
The equality in (\ref{113}) is
achieved for the angular momentum eigenfunctions $F_m(\varphi)=\exp(im\varphi)$.

 Another type of
``entropic'' inequalities was suggested by Deutsch \cite{Deutsch83}, who considered two
noncommuting Hermitian operators, $\hat{A}$ and $\hat{B}$, possessing discrete spectra
and complete orthonormalized sets of eigenvectors $\{|a\rangle\}$ and $\{|b\rangle\}$.
 The entropies $S^A$ and $S^B$ for an arbitrary state $|\psi\rangle$
were defined as follows: 
\[
S^A = -\sum_a |\langle a|\psi\rangle|^2 \ln(|\langle a|\psi\rangle|^2),
\]
\[
S^B = -\sum_b |\langle b|\psi\rangle|^2 \ln(|\langle b|\psi\rangle|^2).
\]
It was proved that these quantities satisfy the
inequality 
\be
S_A(|\psi\rangle) + S_B(|\psi\rangle) \ge
2\ln\left(\frac{2}{1+\mbox{sup}\{|\langle a|b \rangle|\} }\right),
\label{116} 
\ee
where the supremum is taken with respect to
all possible values of the scalar product of  vectors $|a\rangle$ and $|b\rangle$.
 The main advantage of inequality (\ref{116}) over (\ref{unc3}) is that the
right-hand side of (\ref{116}), unlike (\ref{unc3}), {\em does not depend\/} 
on the state $|\psi\rangle$, but is determined (although implicitly) only by the
operators $\hat{A}$ and $\hat{B}$.

If one assumes  the quantities
\be
\Delta_S x = \exp(S_x), \quad \Delta_S p = \exp (S_p)
\label{126}
\ee
as the measures of localization of a particle in the $x-p$ spaces
 \cite{Leipnik59},
inequality (\ref{109}) can be rewritten in a form similar to 
(\ref{unc2}):
\be
\Delta_S x  \Delta_S p \ge (\pi e \hbar)^n.
\label{127}
\ee

One of several examples of inequality (\ref{109}) considered in \cite{Leipnik59}, 
is related to the function
\be
\psi_*(x) = \left\{
\begin{array}{cc}
(2A)^{-1/2} \exp(ik_0 x), & |x-\tilde{x}| \le A
\\
0, & |x-\tilde{x}| > A
\end{array}
\right.
\label{129}
\ee
with the sum $S_x + S_k = \ln(2\pi) +2(1-\gamma)$,
where $\gamma \approx 0.577$ is Euler's constant.
This example  shows a greater effectiveness of inequality (\ref{109}) in
comparison with inequality (\ref{unc2}), which simply has no sense in this
case, because the average value $\langle p^2 \rangle$ does not exist for the function
$\psi_*(x)$. It is worth noting in this connection that the first example
illustrating the relation $\delta x \delta p \gtrsim h$ or $\delta t \delta\omega \gtrsim 1$ 
 in the majority of
textbooks is just the slit diffraction of the de Broglie waves 
in the case of quantum mechanics or light waves  in optics and the Fourier
decomposition of a rectangular pulse in the case of radio
engineering, when the corresponding functions have the form of (\ref{129}).
Reviews of more recent results obtained in the theory of entropic uncertainty relations can be found, e.g., 
in Refs. \cite{Majernik97,Wehner10,Coles17,Hertz19}.

\section{Uncertainty Relations for Mixed States}
\label{sec-mix}

Frequently (e.g., in many text books), the uncertainty relations
are considered for {\em pure\/} quantum states only.
The validity of (\ref{unc2}) and (\ref{unc3}) for {\em mixed\/} quantum states,
described by means of the statistical operator (density matrix) $\hat\rho$
was established for the first time by Mandelstam and Tamm \cite{TM}. 
All inequalities expressing uncertainty relations in terms of variances are valid
both for pure and mixed quantum states. However, 
 the equality in relation (\ref{unc4}) in
general holds only for pure states \cite{DKM80}.
 For example, in the case of an equilibrium state of a harmonic
oscillator with frequency $\omega$ at temperature $T$,
the uncertainty product equals ($k_B$ is the Boltzmann constant)
\be
\sigma_{pp}\sigma_{xx}=\left[\frac{\hbar}{2}
\coth\left(\frac{\hbar\omega}{2k_B T}\right)\right]^2 .
\label{prodT}
\ee
In the high-temperature case, $k_B T \gg \hbar\omega$, the right-hand side
of (\ref{prodT}) is so large, that inequality (\ref{unc2}), in spite of
being absolutely correct, becomes practically useless.
Therefore, the following problem arises naturally: to find
generalizations of inequality (\ref{unc2}) which would contain
 some extra dependence on parameters characterizing
the ``degree of purity'' of a quantum state, in such a way
that generalized
relations could turn into an equality (perhaps, approximate) even for
highly mixed states.


The simplest parameter characterizing the ``purity'' of a quantum state is
$\mu =\mbox{Tr}\hat{\rho}^2$.
Remember that $\hat\rho^2=\hat\rho$ for pure states, so that
$\mu =1$ due to the normalization condition $\mbox{Tr}\hat{\rho }=1$,
whereas 
$\hat\rho^2 \neq \hat\rho$ and $0<\mu<1$ for mixed states.
It is known that for any quantum state described by means of a
{\em Gaussian\/} density matrix (or some 
quasiprobability distribution), the following equality holds
for systems with one degree of freedom \cite{183}:
\be
 \sqrt{\sigma_{pp}\sigma_{xx} - \sigma_{xp}^2}
=\frac{\hbar}{2\mu} .
\label{SR-mu}
\ee
Consequently, one can look for a generalized ``purity bounded uncertainty
relation'' in the form
\begin{equation}
\sqrt {\sigma_{pp}\sigma_{xx} - \sigma_{xp}^2}\ge\frac {\hbar}2\Phi
(\mu ),
\label{79}
\end{equation}
where $\Phi (\mu )$ is a monotonous function of $\mu$,
satisfying the relations $\Phi (1)=1\le \Phi(\mu)\le \mu^{-1}$
for $0<\mu\le 1$.
Actually, it is sufficient to find function $\Phi(\mu)$,
considering a subfamily of states with zero covariance, taking into
account that the quantity
${\cal D} \equiv \sigma_{pp}\sigma_{xx} - \sigma_{xp}^2$
is invariant with respect to arbitrary linear canonical transformations
of operators $\hat{x}$ and $\hat{p}$ \cite{univ00}.

The first explicit expression for the function $\Phi(\mu)$ in the form
$\Phi_B(\mu)=8/(9\mu)$ was found by Bastiaans
\cite{Bast1} in the context of the problem of partially coherent light
beams, where parameter $\mu$ had the meaning of the degree of space
coherence. However, function $\Phi_B(\mu)$ does not satisfy the condition
$\Phi (1)=1$, i.e., using this function one arrives at the inequality
which is weaker than (\ref{unc2}) for pure quantum states.
Besides, the lower limit of the uncertainty
product $4\hbar/(9\mu)$ cannot be achieved for any quantum state.
Actually, $\Phi_B(\mu)$ is an {\em asymptotical form\/} of the
exact expression found in Ref. \cite{183}.
It was shown that minimizing statistical operators are given by finite
diagonal expansions over the first Fock states:
\[
\hat{\rho }_{min}=\sum_{m=0}^M
\rho_m|m\rangle\langle m|,
\quad \rho_m= \frac{2(M+\gamma-m)}{(M+1)(M+2\gamma)},
\]
\[
2\gamma =\left\{\frac {M(M+2)}{3[\mu (M+1)-1]}\right\}^{1/2} -M,
\quad 0\le\gamma\le 1,
\]
\[
\sum\rho_m=1,\qquad\sum\rho_m^2=\mu,
\]
\be
\Phi_{M}(\mu) = 1+M 
- \left\{\frac {M}{3}(M+1)(M+2)
\left[\mu -\frac 1{M+1}\right]\right\}^{1/2}.
\label{90}
\ee

The integer $M$ must satisfy two constraints:
\begin{equation}
\mu^{-1} -1 \le M \le\frac {4-3\mu +\sqrt {16+9\mu^2}}{6\mu}.
\label{89}
\end{equation}
If $\varepsilon \equiv 1-\mu\ll 1$, the only
possible value of $M$ is $M=1$.
In this case, function
$\Phi (\mu )$ takes the form
\begin{equation}
\Phi_1(\mu )=2-\sqrt {2\mu -1} =1+\varepsilon +\varepsilon^2/2+\ldots .
\label{91}
\end{equation}
which is obviously less than
$\mu^{-1}=1+\varepsilon +\varepsilon^2+\ldots$. 

Function (\ref{91}) is well defined for $\mu >1/2$.
However, $\Phi_1(\mu )$
gives the minimal possible value of the product
$\sqrt {\sigma_{pp}\sigma_{xx}}$
only in the interval $1\ge\mu\ge 5/9$.
At point $\mu_2=5/9$ the value $M=2$ becomes admissible
in accordance with (\ref{89}), and the new expression for
$\Phi (\mu )$ emerges:
\begin{equation}
\Phi_2(\mu )=3-\sqrt {8(\mu -1/3)}.
\label{93}
\end{equation}
Both functions, (\ref{91}) and (\ref{93}), coincide at the point $\mu_2$:
\[
\Phi_1(5/9)=\Phi_2(5/9)=5/3.
\]
Moreover, the first derivatives of $\Phi_1(\mu )$ and $\Phi_2(\mu )$ 
 coincide at $\mu=\mu_2$.
 But we have $\Phi_1(\mu )>\Phi_2(\mu )$ for $\mu <5/9$.
It is easy to verify that for the given value of purity $\mu$,
the minimal value of function (\ref{90}) is achieved for the maximal
admissible value of integer $M$, because the derivative of function
(\ref{90}) with respect to $M$ equals $-\infty$ at $M=\mu^{-1}-1$,
which means that this function decreases with increase of $M$.

Thus we arrive at the set of functions
$\Phi_k(\mu )$,
representing the minimizing function
$\Phi (\mu )$ in the intervals $\mu_k\ge\mu\ge\mu_{k+1}$,
where the boundary points $\mu_k$
are determined from the condition that
(\ref{89}) becomes an equality for $M=k$:
\begin{equation}
\mu_k=\frac {2(2k+1)}{3k(k+1)}=
\frac 4{3k}\left[1-\frac 1{2(k+1)}\right].
\label{94}
\end{equation}
In particular,
\[
\mu_3=\frac 7{18},\quad\mu_4=\frac3{10},
\]
\[
\Phi_3(\mu )=4-\sqrt {20(\mu -1/4)},\quad
\Phi_4(\mu )=5-\sqrt {40(\mu -1/5)},
\]
\begin{equation}
\Phi (\mu_k)=\frac {1+2k}3=\frac {4+\sqrt {16+9\mu_k^2}}
{9\mu_k}.
\label{95}
\end{equation}
The first derivatives of functions $\Phi_k(\mu )$ and $\Phi_{k+1}(\mu )$
coincide at the boundary points $\mu_{k+1}$. But their derivatives
of higher orders are different at these points.

Since explicit analytical forms of function $\Phi (\mu )$
are different for different segments of the interval $0<\mu\le 1$,
it can be convenient to use a simple interpolation
expression, replacing an integer $M$ in
(\ref{90}) by its maximal admissible value (\ref{89}) (even if this value
is not integral):
\be
\tilde{\Phi }(\mu )=\frac {4+\sqrt {16+9\mu^2}}{9\mu}.
\label{Phitil}
\ee
The functions $\tilde{\Phi }(\mu )$ and ${\Phi }(\mu )$ coincide at
points $\mu_k$. 
One has $\tilde{\Phi} (\mu )>{\Phi} (\mu )$
immediately to the right from these points,
whereas $\tilde{\Phi} (\mu )<{\Phi} (\mu )$
immediately to the left of these points.
However, the difference between the exact and approximate values does not
exceed $0.02$ even for the values of $\mu$ close to unity. Moreover,
for $\mu\to 0$ this difference becomes less than $\mu^2/64$.
For $\mu\ll 1$, the following asymptotical formula holds:
\begin{equation}
\tilde{\Phi }(\mu )=\frac 8{9\mu}\left(1+\frac 9{64}
\mu^2+\ldots\right),
\label{97}
\end{equation}
and $|{\Phi} (\mu )-8/(9\mu)|<0,01$ for $\mu\le 0,25\approx \mu_5$.

\section{``Local'' Uncertainty Relations}

The inequalities considered in the preceding sections characterize the behavior of
the distributions $|\psi(x)|^2$ and $|\vf(p)|^2$ as a whole. 
Several authors studied the relations between local properties of the
function $\psi(x)$ and its Fourier transform $\vf(p)$. 
To understand the significance of the so-called ``local'' uncertainty relations, 
let us suppose that, for example, the function $\vf(p)$ possesses a
sharp maximum, so that the value of $\Delta p$ is small. Using re1ation (\ref{dxp}), one can
conclude that the value of $\Delta x$ must be large. However, 
the large value of $\Delta x$ would not contradict the assumption that the
function $\psi(x)$ has two sharp ``humps'' located at a great distance from each other. 
The essence of the ``local uncertainty relations'' consists in the statement that 
in reality not only the value of $\Delta x$ is great when $\Delta p$ is small,
but the probability density $|\psi(x)|^2$ is always small (so that two sharp
``humps'' cannot exist). The exact formulation is as follows \cite{Faris}. 

Let $a$ be an arbitrary real parameter and $b$ any positive number. Then, the
probability of finding the particle in the domain $|x - a| \le b$ in the one-dimensional
case satisfies the inequality
\be
\mbox{Prob}\{|x-a|\le b\} \le 2b\Delta p/\hbar, \qquad \Delta p=\sqrt{\sigma_p}.
\label{(206)}
\ee
To prove (\ref{(206)}), let us consider the
 inequality (\ref{156}) with $n = 1$, $\nu = \nu_1 = 0$, and the function
$\vf(x)=(2/{\pi})\arctan\left[(x-b)/{\vep}\right]$.
Since $|\vf(x)|\le 1$, formula (\ref{156}) leads to the inequality
\be
\hbar\langle \delta_{\vep}(x-b)\rangle \le \sqrt{\langle \hat{p}^2\rangle}, \qquad
\delta_{\vep}(t) \equiv \frac{\vep}{\pi\left(t^2 +\vep^2\right)}.
\label{(207)}
\ee
Note that the function $\delta_{\vep}(t)$ is nothing but an approximation of the delta-function
for $\vep\to 0$. Performing the limit transition
$\vep \to 0$ in (\ref{(207)}) and replacing $\hat{p}$ by $\hat{p} -\langle \hat{p}\rangle$,
one can obtain the  inequality (it was also derived in \cite{Baum})
\be
|\psi(y)|^2 \le \Delta p/\hbar, \qquad -\infty < y < \infty,
\label{(208)}
\ee
which is equivalent to (\ref{(206)}). Thus, the variance of the momentum restricts
the maximum value of the wave function in the coordinate representation
(and vice versa). Another interpretation of (\ref{(208)}) is also  possible:
\be
\Delta p \ge \hbar \mbox{max}\left\{|\psi(y)|^2\right\},
\label{(208a)}
\ee
i.e., the maximum value of the (normalized) coordinate wave function gives a lower
boundary for the variance of the momentum. Moreover, the following generalization
of relation (\ref{dxp}) is the consequence of (\ref{(208a)}) for the normalized wave functions:
\be
\Delta p \Delta x \ge \hbar^2 \mbox{max}\left\{|\psi(x)|^2\right\}\mbox{max}\left\{|\vf(p)|^2\right\}.
\label{(208b)}
\ee

 A generalization of inequality (\ref{(208)}) was
obtained in \cite{LevLeb}. In that paper, inequality (\ref{unc3}) was applied to the operators
$\hat{A} = \hat{p}$ and $\hat{B} = F(x) = \theta(x - x_0)$, where $\theta(x)$ is the Heaviside function. 
Then,
$\langle F'\rangle = \langle\delta(x - x_0)\rangle = |\psi(x_0)|^2$, whereas
\[
\langle F^2\rangle =\langle F\rangle =\int_{x_0}^{\infty} |\psi(x)|^2 dx = 1-{\cal P}(x_0),
\]
where ${\cal P}(x_0)$ is the probability for localizing the particle in the interval $(-\infty,x_0)$. 
The result is the inequality
\be
|\psi(x)|^2 \le \frac{2\Delta p}{\hbar}\sqrt{{\cal P}(x)[1-{\cal P}(x)]}\le \frac{\Delta p}{\hbar}
\label{(213)}
\ee
(since ${\cal P}(1-{\cal P}) \le 1/4$).
 This relation turns into an equality for the function
 $\psi(x)=a^{-1/2}\exp(-|x|/a)$
at the point $x = 0$. Taking into account that $|\psi(x)|^2=d{\cal P}/dx$
 and integrating both sides of inequality (\ref{(213)}), one obtains the relation
 (assuming that $ x \ge x_0$)
\be
\mbox{arcsin} \sqrt{{\cal P}(x)} - \mbox{arcsin} \sqrt{{\cal P}(x_0)}
\le \Delta p(x-x_0)/\hbar, 
\label{(214)}
\ee
which can be written as
\be
{\cal P}(x) \le \sin^2\left[\mbox{arcsin} \sqrt{{\cal P}(x_0)} +\Delta p(x-x_0)/\hbar\right],
\label{(215)}
\ee
provided $\mbox{arcsin} \sqrt{{\cal P}(x_0)} +\Delta p(x-x_0)/\hbar \le \pi/2$.

The following ``quadratic local uncertainty relation'' was found in the multidimensional 
case ($n \ge 3$) in  \cite{Faris}:
\be
\mbox{Prob} \left\{| {\bf x} - {\bf a}| \le b\right\} \le \frac{4}{(n-2)^2}
\left(\frac{b\Delta p}{\hbar}\right)^2, 
\label{(219)}
\ee
\[
(\Delta p)^2 =\left\langle\left(\hat{\bf p} -\langle\hat{\bf p}\rangle\right)^2\right\rangle.
\]
To prove (\ref{(219)}), let us introduce the characteristic function of the set $|{\bf x}| \le b$:
\[
\chi({\bf x}) =\left\{
\begin{array}{ll}
1, & |{\bf x}| \le b,
\\
0, & |{\bf x}| >b,
\end{array} 
\right.
\]
\[
\int \chi({\bf x})|\psi({\bf x})|^2 d{\bf x} =  \mbox{Prob} \left\{| {\bf x} | \le b\right\}.
\]
It is obvious that $\chi({\bf x})b^{-2} \le r^{-2}$.
 Averaging this relation over the distribution $|\psi({\bf x})|^2$
 and taking into account inequality (\ref{161}), one obtains inequality (\ref{(219)})
after the substitutions ${\bf x} \to {\bf x} - {\bf a}$ and 
$\hat{\bf p} \to \hat{\bf p} -\langle\hat{\bf p}\rangle$.  It can be
strengthened for $3 \le n \le 5$, if one considers inequality (\ref{156}) with $\nu=\nu_2 \le 0$ and the negative
function
$\vf (r) = - (\pi\hbar/2b) \cot (\pi r/2b) \chi (r) $.
Then,
\[
\hbar\langle \frac{\partial \vf}{\partial r}\rangle \le \hbar\langle \frac{\partial \vf}{\partial r}
-(n-3)\frac{\vf}{r}\rangle \le 2\sqrt{\langle \vf^2\rangle\langle\hat{\bf p}^2\rangle}
\le \langle \vf^2\rangle +\langle\hat{\bf p}^2\rangle,
\]
i.e.,
$\langle\hat{\bf p}^2\rangle \ge \langle \hbar \partial \vf/\partial r - \vf^2\rangle
=(\pi\hbar/2b)^2\langle\chi({\bf x})\rangle $,
from which it follows that
\be
\mbox{Prob} \left\{| {\bf x} - {\bf a}| \le b\right\} \le (2b \Delta p/\pi\hbar)^2.
\label{(220)}
\ee

Note that for proving local inequalities (\ref{(206)}), (\ref{(219)}) or (\ref{(220)}) the actual
meaning of the averaging procedure denoted by the symbol $\langle\ldots\rangle$ is quite insignificant.
Consequently, these relations are valid for mixed quantum states
described in terms of a density matrix as well as for pure states. In particular,
relation (\ref{(208)}) can be generalized as follows:
\be
\rho (x, x) \le \Delta p/\hbar, \qquad - \infty < x < \infty. 
\label{(221)}
\ee
It is worth noting that  no ``quadratic'' inequality like (\ref{(220)}) exists
in the two-dimensional case: this was shown in study \cite{Faris}.

\section{``Total Width'' and ``Mean Peak Width''}
\label{sec-HilUf}

Several characteristics of ``uncertainties'' different from variances and entropies 
have been introduced by Hilgevoord and Uffink in 1980s.
Two main notions 
are the ``total width'' (TW) of the wave function and its ``mean peak width'' (MPW).
The total width of function $\psi(x)$ was defined as
the smallest value $Q$ for which the probability of finding the particle in the
interval $(x_0 -Q/2,x_0+Q/2)$  equals a given number 
$\alpha \le 1$:
\[
Q_{\alpha}(x_0 \vert \psi) = \mbox{min}\left\{X: \; \int_{x_0-X/2}^{x_0+X/2} |\psi(x)|^2 dx = \alpha\right\}.
\]
The mean peak width $q_{\beta}(\psi)$ was
defined as the smallest positive number  for which the 
absolute value of the coordinate correlation function 
equals a given number $\beta$:
\be
q_{\beta} = \mbox{min}\left\{q: \; |R_{\psi}(q)| =\beta \right\},
\label{(228)}
\ee
\be
R_{\psi}(q) =\int \psi^*(x)\psi(x-q) dx = \int |\vf(k)|^2 e^{ikq} dk .
\label{(229)}
\ee
The notions of TW and MPW are useful when
variances do not exist. For example, if
\[
\psi(x)=\sqrt{\frac{a}{\pi}} \, \frac{\sin(ax)}{ax},
\] 
then, $Q \sim 1/a$, while $\sigma_x=\infty$. Besides, they are useful in the case of
``two-hump'' functions, 
when the total width characterizes the distance
 between ``humps'' while the MPW characterizes the average width of ``humps''. 
Various relations between the quantities $Q_{\alpha}$, $q_{\beta}$ and variances were obtained  in  studies 
\cite{Price,UffHill85}.  
The local  uncertainty  relation (\ref{(206)}) leads immediately to the inequalities 
(for arbitrary values $x_0$ and $p_0$) \cite{Price}
\be
Q_{\alpha}(x_0 \vert \psi) \ge \hbar\alpha/\Delta p, \quad
Q_{\alpha}(p_0 \vert \vf) \ge \hbar\alpha/\Delta x.
\label{(231)}
\ee 
Due to the Chebyshev inequality (see, for example, \cite{Feller})
\be
\mbox{Prob} \left\{| x - \langle x\rangle| \ge b\right\} \le \sigma_x/b^2 ,
\label{(216)}
\ee
 the quantity $Q_{\alpha}(\psi)$  (with $x_0 = \langle x \rangle$)
is bounded from above by the variance of the coordinate 
\cite{Price,HilUff83,UfHil84}, and the quantity $Q_{\alpha}(\vf)$  is bounded by the variance of the momentum:
\be
Q_{\alpha}(\psi) \le \frac{2\Delta x}{\sqrt{1-\alpha}}, 
\quad Q_{\alpha}(\vf) \le \frac{2\Delta p}{\sqrt{1-\alpha}}.
\label{(232)}
\ee
The mean peak width is also bounded from below by the quantity $(\Delta p)^{-1}$ \cite{UfHil85}: 
\be
q_{\beta}(\psi) > \frac{\hbar}{\Delta p}\sqrt{2(1-\beta)}. 
\label{(233)}
\ee 
Another estimate was given in study \cite{Price}:
\be
q_{\beta}(\psi) > \frac{2\hbar}{\Delta p}(1-\beta^2).
\label{(234)}
\ee
The estimate (\ref{(233)}) is stronger than (\ref{(234)}) for $\beta > 0.855$. 

An  upper limit for the mean peak width was found in study \cite{LevLeb} on the basis of the inequality
 ($-\infty < k < \infty$)
\be
|R_{\vf}(k)|^2 \equiv \left|\langle e^{ikx}\rangle\right|^2 \le
\left[1 +\hbar^2 k^2/(4\sigma_p)\right]^{-1},
\label{(235)}
\ee
which is a consequence of inequalities
\[
\sigma_p\left[\langle\cos^2(kx)\rangle - \langle\cos(kx)\rangle^2\right]
\ge \frac14 k^2\hbar^2 \langle\sin(kx)\rangle^2,
\]
\[
\sigma_p\left[\langle\sin^2(kx)\rangle - \langle\sin(kx)\rangle^2\right]
\ge \frac14 k^2\hbar^2 \langle\cos(kx)\rangle^2,
\]
following, in turn, from the general inequality (\ref{unc3}) under the choice $\hat{A} = \hat{p}$
and $\hat{B}=\cos(kx)$ or $\sin(kx)$.
 Taking into account the definition (\ref{(228)}), one can see
that consequencies of (\ref{(235)}) are the inequalities
\be
q_{\beta}(\psi) \le \frac{2\Delta x}{\beta}\sqrt{1-\beta^2}, \quad
q_{\beta}(\vf) \le \frac{2\Delta p}{\beta\hbar}\sqrt{1-\beta^2}. 
\label{(237)}
\ee
A consequence of (\ref{(237)}) and a weakened version of (\ref{(233)}),
\be
q_{\beta}(\psi) > (\hbar/\Delta p)\sqrt{1-\beta^2},
\label{(238)}
\ee
is the inequality $\Delta x \Delta p \ge \beta\hbar/2$. Since it must hold for any value
$\beta\le 1$, we arrive again at inequality (\ref{dxp}).

The uncertainty relation in terms of TW and MPW has the form \cite{UffHill85}
\be
Q_{\alpha}(x_0|\psi)q_{\beta}(\vf) \ge 2\mbox{arccos}\left[(1 +\beta -\alpha)/{\alpha}\right],
\label{(242)}
\ee
provided $\beta \le 2\alpha -1$.
Examples of functions giving the equality in (\ref{(242)}) were found in paper \cite{UffHill85}. 

An inequality which is to some extent the opposite of (\ref{(242)}), 
was derived in study \cite{UfHil84} for $\beta^2 \ge\alpha$:
\be
\mbox{min}_{x_0}\left\{Q_{\alpha}(x_0|\psi)\right\}
q_{\beta}(\vf) \le \frac{2(1-\alpha)}{\beta^2 -\alpha}.
\label{(243)}
\ee
The usual uncertainty relations or inequality (\ref{(242)})  assert that if
one of two functions, $\psi$ or $\vf$, is narrow, then the other one ought to be wide.
The significance of inequality (\ref{(243)})  consists in the assertion that if
one of the two functions is wide, then the other {\em ought to be narrow}.

Inequalities $0 \le \beta \le 2\alpha - 1$ and $1 \ge \beta^2 \ge \alpha \ge 0$ are 
incompatible for arbitrary values of $\alpha$ and $\beta$ (excluding the case of $\alpha =\beta=1$). 
Thus, for any pair of numbers $\alpha$ and $\beta$, the product $Q_{\alpha}(\psi)q_{\beta}(\vf)$
 can be bounded either from above [according to inequality (\ref{(242)})] 
 or from below [as in relation (\ref{(243)})],
 but not from two sides simultaneously.  

The concepts of TW or MPW were used, for example, in studies \cite{UffHill85,UfHil85,HilUf85} 
 for the interpretation of the results of experiments on testing uncertainty relations with the aid of a neutron interferometer, since in such experiments one measures only the coordinate correlation function $R_{\psi}(q)$; 
in addition, the total width and the mean peak width can be determined from the experimental curve of the momentum distribution more reliably than the variance.

\section{Phase and Angle}
\label{sec-phase}

A similarity between the formula $\hat{L}_z = -i\hbar\partial/\partial \vf$ for the  component 
of the angular momentum operator and the formula $\hat{p} = -i\hbar \partial/\partial x$ for the linear momentum operator 
suggests an idea of the existence of the inequality $\Delta L_z \Delta \vf \ge \hbar/2$. 
However, this relation is incorrect (although it can be met in some simplified textbooks).
The well known counterexamples are the eigenstates of $\hat{L}_z$ with
$\psi(\vf) \sim \exp(im\vf)$ and $\Delta L_z =0$. 
The origin of difficulties is the non-Hermiticity  of the operator $\hat\vf$ understood as the simple multiplication
operator: if $\psi(\vf)$ is a periodic function, then $\vf\psi(\vf)$ is obviously non-periodic.
The problem of correct description of phase or angle in quantum mechanics was discussed in many
publications: see a few examples \cite{Carr68,Jackiw,Lev76,Breit85,PB89,Lynch95,Perina-ang,Kastrup06,Gaz16}.
One of possible solutions is to use the triple of operators \cite{Carr68}
\be
\hat{L} \equiv \hat{L}_z= -i\hbar\partial/\partial\varphi, \quad
\hat{C} =\cos\varphi, \quad \hat{S} =\sin\varphi,
\label{def-LCS}
\ee
satisfying the commutation relations 
\[
\left[\hat{L}, \hat{S}\right]= -ih \hat{C}, \quad \left[\hat{L}, \hat{C}\right]= ih \hat{S}, \quad 
\left[\hat{C}, \hat{S}\right]= 0.
\]
Inequalities containing operators
$\hat{C}$ and $\hat{S}$  on an equal footing were given, e.g., in Refs. \cite{Carr68,Jackiw}.
For example,
\be
\overline{L^2}\left(\overline{C^2}+\overline{S^2}\right)
\ge \frac{\hbar^2}{4}\left[\left(\overline{C}\right)^2
+ \left(\overline{S}\right)^2\right].
\label{unc31}
\ee
Since $\overline{C^2}+\overline{S^2} \equiv 1 - \left(\overline{C}\right)^2 - \left(\overline{S}\right)^2$,
one can write
\be
\left(\overline{C}\right)^2 + \left(\overline{S}\right)^2 \le \frac{\overline{L^2}}{\overline{L^2} +\hbar^2/4}.
\label{unc31-new}
\ee
Inequality (\ref{prod3}) yields
\be
\overline{L^2}\cdot\overline{C^2}\cdot\overline{S^2} \ge \frac49\hbar^2\left[
\overline{C^2}\left(\overline{C}\right)^2 + \overline{S^2}\left(\overline{S}\right)^2\right].
\label{LCS49}
\ee

 However, it can be interesting nonetheless to obtain an inequality just for the product $\Delta \vf \Delta L_z$. 
This can be done if one writes the inequality 
\[
\int_{\vf_0}^{\vf_0 + 2\pi} \left|\alpha(\vf -\langle \vf \rangle)\psi +\hbar \partial \psi/\partial \vf
-i\langle L_z\rangle \psi \right|^2 d\vf \ge 0
\]
with $\mbox{Im}(\alpha) =0$,
integrating by parts the left-hand side and remembering that the terms 
$\left.\vf |\psi(\vf)|^2\right|_{\vf_0}^{\vf_0 + 2\pi}$  do not vanish
(contrary to the $xp$-case, where $\left. x |\psi(x)|^2\right|_{-\infty}^{\infty}=0$). 
Using the condition of non-negativeness of the second-degree polynomial with respect to the parameter $\alpha$,
one obtains the correct inequality 
%
\be
\sigma_{\vf}\sigma_{L} \ge \left(\hbar^2/4\right)\left(1 -2\pi|\psi(\vf_0)|^2\right)^2, 
\label{(259)}
\ee
where $\vf_0$ is the left boundary of the interval $(\vf_0, \vf_0 +2\pi)$ 
chosen as the domain  of the unambiguous definition of the angle variable
[so that $\psi(\vf_0) = \psi(\vf_0 +2\pi)$]. 
The right-hand side of (\ref{(259)}) equals zero 
for the angular momentum eigenfunctions $\psi_m(\vf) = (2\pi)^{-1/2}\exp(im\vf)$. 
 In the general case, 
the inequality  $\Delta \vf \Delta L \ge \hbar/2$
  is possible only for functions $\psi(\vf)$ satisfying the restriction  $\psi(\vf_0) = 0$.
Inequality (\ref{(259)}) could also be derived from the general inequality (\ref{unc3}) 
with the aid of the correct commutation relation for the phase operator \cite{Carr68}
\be
\left[\hat\vf, \hat{L}\right] =i\hbar\left\{1 - 2\pi \sum_{n=-\infty}^{\infty}\delta(\vf -[2n+1]\pi)\right\}.
\label{(260)}
\ee
For other measures of the phase (angle) uncertainty and related inequalities one can consult, e.g.,
Refs. \cite{Trif03circ,Hradil06,Hradil10,Prz16}.

\section{Time-Energy Uncertainty Relations}
\label{sec-TE}

The energy--time uncertainty relation (ETUR)
\be
\Delta E \Delta T  \gtrsim h
\label{ETh}
\ee
is one of the most famous and at the same time most controversial formulas of quantum theory.
It was introduced by Heisenberg \cite{Heis} together with his coordinate--momentum
uncertainty relation 
\be
\Delta x \Delta p  \gtrsim h.
\label{xph}
\ee
The importance of both relations for the interpretation of quantum mechanics was emphasized by Bohr \cite{Bohr28}.
However, the further destiny of relations (\ref{ETh}) and (\ref{xph}) turned out quite different.
A strict formulation of relation (\ref{xph}) was found almost immediately by Kennard \cite{Kennard} 
in the form of inequality (\ref{dxp}),
where $\Delta x$ and $\Delta p$ are well-defined quantities, namely mean-square deviations. 

On the contrary, the physical and mathematical meanings of inequality (\ref{ETh}) appeared to be much less clear than that
of (\ref{dxp}) (and less clear than Heisenberg, Bohr and other creators of quantum mechanics thought initially). 
The main reason is that, in fact, there are several quite different physical problems where relations like
(\ref{ETh}) can arise, and in each concrete case the meaning of the quantities
standing on the left-hand side proves to be different. This was demonstrated for the first time 
 by Mandelstam and Tamm \cite{TM} 
and by Fock and Krylov \cite{FoKry}, and many authors arrived at the same conclusions later
\cite{AharBohm61,Allcock,Bauer,Busch90-1}.

\subsection{Mandelstam--Tamm inequalities and their application to the problem of decay} 
\label{sec-MT}

The first rigorous formulations of relation (\ref{ETh}) was given by Mandelstam and Tamm \cite{TM}. 
Choosing $\hat{B}=\hat{H}$ (where $\hat{H}$ is the system
Hamiltonian) and remembering that operator $(i/\hbar)[\hat{H}, \hat{A}]$ is the operator of
the rate of change of the quantity $A$, i.e.,  
$(i/\hbar)[\hat{H}, \hat{A}]=d\hat{ A}/dt$  (provided operator $\hat{A}$
does not depend on time explicitly), they transformed inequality (\ref{unc3})
to the form
\be
\Delta E \Delta t_A \ge \hbar/2,
\label{(299)}
\ee
where $\Delta E \equiv \Delta H$ and
\be
 \Delta t_A \equiv \Delta A/|\langle d\hat{ A}/dt\rangle|.
\ee
The meaning of inequality (\ref{(299)}) is that it yields an estimate of the time interval 
required for a significant change in the average value of observable $A$:
by an amount of the order of the mean squared variation. 
It results in the statement that  ``a dynamical quantity cannot change,
remaining always dispersionless'' \cite{TM}.

Relation (\ref{(299)}) may seem artificial at first glance, since operator $\hat{A}$ may be quite arbitrary.
However, there exists at least one important specific choice of this operator. 
Let us consider, following  \cite{TM}, the projector $\hat{L} = |\psi(0)\rangle\langle \psi(0)|$ 
on some initial quantum state $|\psi(0)\rangle$. 
Since $\hat{L}$ is the projection operator, $\hat{L}=\hat{L}^2$ and
\[
\Delta L \equiv \sqrt{\overline{(L^2)} -(\overline{L})^2} =
\sqrt{\overline{L} -(\overline{L})^2}.
\]
Then, inequality (\ref{(299)}) assumes the form
\be
-\frac{\hbar}{2}\frac{d\overline{L}}{dt} =\frac{\hbar}{2}\left|\frac{d\overline{L}}{dt}\right|
\le \Delta E \sqrt{\overline{L} -(\overline{L})^2}.
\label{eq-L}
\ee
Integrating (\ref{eq-L}) with account of the initial condition $\overline{L}(0)=1$,
Mandelstam and Tamm obtained relations
\begin{equation}
\Delta Et/\hbar\ge\pi /2-\arcsin\sqrt {Q(t)}, \quad 
0\le t<\infty ,
\label{330a}
\end{equation}
\begin{equation}
Q(t)\ge\cos^2(\Delta Et/\hbar ), \quad 0\le t\le\pi
\hbar /2\Delta E,
\label{330b}
\end{equation}
where
\be
Q(t) \equiv \overline{L}(t) \equiv \langle\psi(t)|\hat{L}|\psi(t)\rangle =
|\langle \psi(0)|\psi(t)\rangle |^2. 
\label{defQ}
\ee
One may interpret function $Q(t)$ as the probability to remain in the
initial state $|\psi(0)\rangle$. In this case, it seems natural to define the half-decay period 
$T_{1/2}$ by means of the relation $Q\left(T_{1/2}\right)=1/2$.
 Then, Eq.  (\ref{330b}) results in the inequality
\begin{equation}
T_{1/2}\Delta E\ge\pi\hbar /4\approx 0.785\hbar
\label{332}
\end{equation}
with well defined quantities $T_{1/2}$ and $\Delta E$.

An immediate important consequence of inequality  (\ref{330b})
is the impossibility of strictly exponential decay
\begin{equation}
Q(t)=\exp(-t/\tau )
\label{327}
\end{equation}
for realistic physical systems with a finite energy dispersion $\Delta E$ \cite{TM,FoKry,Fleming73,Fonda78}.
Indeed, it follows from (\ref{330b}) that the law (\ref{327})
can be realized only approximately and for sufficiently big values of time satisfying the inequality
\be
t> \tau\ln\left[1+ \hbar^2/(2\tau\Delta E)^2\right]. 
\label{t-tau}
\ee
Only a parabolic time dependence of $Q(t)$ is permitted for short times,
due to the Taylor expansion at $t\to 0$:
\beqnn
Q(t) &=& \left|\langle \psi_0|\exp\left(-\frac{i\hat{H}t}{\hbar}\right)|\psi_0\rangle\right|^2  
 \approx \left| 1 -\frac{it \langle\hat{H}\rangle}{\hbar}
-\frac{t^2\langle\hat{H}^2\rangle}{2\hbar^2} \right|^2  \nonumber
\\ &\approx&
1-\frac{t^2}{\hbar^2}\left(\langle\hat{H}^2\rangle -\langle\hat{H}\rangle^2\right)
+ \cdots .
\eeqnn

Generalizations of relations (\ref{330a}) and (\ref{330b}) to the case of 
time-dependent Hamiltonians were obtained
in \cite{Pfeifer95}.
Since inequality (\ref{(299)}) becomes meaningless in the case of $\Delta E=\infty$,
 Mandelstam and Tamm wrote in \cite{TM} that ``it would be desirable to find a more general relation
of the same type as (\ref{(299)})''. Several concrete generalizations
are demonstrated in the subsequent subsections.

\subsection{Decay laws and spectral distributions}

The relations between the decay law $Q(t)$ and the energy spectrum of the system
were established by Krylov and Fock \cite{FoKry}.
Suppose that we know the decomposition of vector $|\psi(0)\rangle$ over the
energy eigenstates:
\be
|\psi (0)\rangle =\int a(E)|E\rangle\,dE, \quad \langle E'|E\rangle =\delta(E-E').
\ee
Then, 
\be
|\psi (t)\rangle =\int a(E)\exp\left(-iEt/\hbar\right)|E\rangle\,dE ,
\label{psitE}
\ee
and the nondecay amplitude $\chi(t)=\langle\psi(0)|\psi(t)\rangle$
(called `integrity amplitude' in \cite{Fleming73})
 can be expressed as the
Fourier transform of the positive energy distribution function $P(E)=|a(E)|^2$:
\begin{equation}
\chi (t) \equiv \langle\psi(0)|\psi(t)\rangle =\int P(E)\exp\left(-iEt/\hbar\right)\,dE.
\label{326}
\end{equation}
Its consequence is the identity
$\int P(E)dE=1$.
The probability of finding the system in the initial state at time instant $t$ equals 
$Q(t)=|\chi (t)|^2$.
The energy variance can be calculated as
\be
(\Delta E)^2=\int\left(E-\bar E\right)^2P(E)\,dE
\label{varE}
\ee
with
$\bar {E}=\int EP(E)\,dE$.
 Fock and Krylov \cite{FoKry} proved that the
necessary and sufficient condition of decay (when $Q(t)\to 0$ for $t\to\infty$)
is the continuity of the integral energy distribution function
$\tilde{P}(E)= \int^{E} P(\vep)d\vep$.
This means, in particular, that the energy spectrum must be continuous, in
order that function $P(E)$ would not contain terms like $\delta(E - E_0)$ corresponding
to discrete energy levels. 

It is known that the exponential decay law (\ref{327}) corresponds to the Lorentzian
energy distribution 
\begin{equation}
P(E)=\frac {\Gamma /2\pi}{\left(E-E_0\right)^2+\Gamma^2/4}.
\label{328}
\end{equation}
The peak width $\Gamma$, defined as the length of the energy interval
where $P(E) \ge P(0)/2$, is related to the lifetime $\tau$ by the equality
\be
\tau\Gamma = \hbar.
\label{(329)}
\ee
However, distribution (\ref{328}) is an idealization (although very good one in many practical cases), 
because it results in the relation $\Delta E=\infty$. Another drawback of distribution (\ref{328}) is
that it implies that the energy spectrum stretches from $-\infty$ to $\infty$
(only under this assumption integral (\ref{326}) yields the exponential function of time).
But the energy of real physical systems is limited from below, and this fact leads to violations of the exponential decay law
for $t\to\infty$, when $Q(t)\sim t^{-\beta}$ with some constant $\beta$ depending on the concrete form of 
the energy spectrum \cite{Khalfin} (see \cite{Fonda78,Newton61,Terentiev72,Peres80,Garcia96,Campo11} for later discussions and reviews).
Moreover, an oscillatory decay is also possible \cite{Peshkin14}.

A simple example of the `decay' that never has the exponential form was given by Bhattacharyya \cite{Bhat83}:
a freely expanding Gaussian wave packet in one dimension has the energy distribution function
\be
P(E) = \left(\sqrt2 \pi  E \Delta E\right)^{-1/2}\exp\left(-E/\sqrt{2}\Delta E\right)
\ee
and the nondecay probability
\be
Q(t) =\left(1 +2(\Delta E)^2 t^2/\hbar^2\right)^{-1/2}.
\ee
In this case, $T_{1/2}\Delta E =\hbar\sqrt{3/2}$.

One of consequences of formula (\ref{326}) was obtained by Luo \cite{Luo-jpa05}:
\be
Q(t) \le \frac12\left[1 +\sqrt{Q(2t)}\right].
\label{Luo}
\ee
Inequality (\ref{Luo}) forbids many decay laws that one could imagine. For example, it clearly
forbids the instantaneous decay, such that $Q(t)=1$ for $t<t_*$ and $Q(t)=0$ for $t>t_*$.
For the recent analysis of the Mandelstam--Tamm UR one can consult \cite{Gray05}.

\subsection{Different decay times of unstable systems}
\label{sec-decay2}

Besides the half-decay time $T_{1/2}$, 
many other definitions of the decay time are possible. 
Fleming \cite{Fleming73} suggested to use the quantity
\begin{equation}
\tau_0=\int_0^{\infty}Q(t)\,dt
\label{333}
\end{equation}
where the non-decay probability $Q(t)$ was defined by equation (\ref{defQ}).
Definition (\ref{333}) gives the lifetime $\tau_0=\tau$ for the exponential function (\ref{327}). 
Taking into account inequality (\ref{330b}) and writing $z= {\pi\hbar }/(2\Delta E)$,
one can obtain the following lower bound  for $\tau_0$ \cite{Fleming73}:
\begin{equation}
\tau_0\ge\int_0^{z}Q(t)\,dt\ge\int_
0^{z}\cos^2\left({\Delta Et}/{\hbar}\right)\,dt=\frac {\pi\hbar}{4\Delta E}.
\label{334}
\end{equation}

The most strong inequality with an achievable lower bound was derived in paper \cite{Gisl85}:
\begin{equation}
\tau_0\Delta E\ge\frac {3\pi}{5\sqrt {5}}\hbar\approx 
0.843\hbar.
\label{342}
\end{equation}
The equality sign in (\ref{342}) is achieved for the distribution 
\be
P_*(E) = \frac{\sqrt{45}}{20 \Delta E}\left(1-\frac{E^2}{5(\Delta E)^2}\right)
\label{(339)}
\ee
(and $P_*(E) = 0$ for $|E| \ge \sqrt{5}\Delta E$), 
where the energy uncertainty $\Delta E$ can be arbitrary positive number 
(provided the energy scale
is shifted in such a way that $\langle E \rangle=0$).
The corresponding  `nondecay probability' can be calculated with the aid of formulas
(\ref{defQ}) and (\ref{326}). It has the form
\begin{equation}
Q_*(t)= 9\left(\sin z- z\cos z\right)^2/{z^6},
\label{343}
\end{equation}
where
$ z(t)= \sqrt{5} t\Delta E/\hbar = 3\pi {t}/(5\tau_0)$.

Two other examples of energy distributions possessing products $\tau_0\Delta E$ close
to the minimal possible value (\ref{342}) were given in Ref. \cite{Gisl85}. The first of
them is the Gaussian distribution
\begin{equation}
Q(t)=\exp\left[-\pi\left(t/2\tau_0\right)^2\right]=\exp\left[-\left(t\Delta E/\hbar\right)^2\right] ,  
\label{344}
\end{equation}
\be
P(E)=\left(\Delta E \sqrt{2\pi}\right)^{-1}\exp\left[-\frac12\left(E/\Delta E\right)^2\right],
\label{344P}
\ee
\be
\tau_0\Delta E= \hbar \sqrt{\pi}/2\approx 0.886 \hbar.
\label{344Et}
\ee
One can check that the nondecay probability (\ref{344}) satisfies the Luo inequality (\ref{Luo}).
The second example is the stepwise energy distribution:
\begin{equation}
P(E)=\left\{\begin{array}{cc}
\sqrt{3}/(6\Delta E), &|E|\le \sqrt{3}\Delta E\\
0, &|E|> \sqrt{3}\Delta E\end{array}
\right. ,
\label{345}
\end{equation}
\be
Q(t)=\left[\frac {\sin\left(\sqrt{3} t\Delta E/\hbar\right)}{\sqrt{3} t\Delta E/\hbar}\right]^2, 
\label{345a}
\ee
\be
\tau_0\Delta E=\frac {\pi \hbar}{2\sqrt {3}}\approx 0.907 \hbar.
\ee
Mathematical aspects of the time decay problem (time asymmetry) were discussed in Refs.
 \cite{Bohm81,Civi04,Marchetti13}.

\subsection{Modifying definitions of the decay times and energy spread}
\label{subsec-modifdecaytimes}

The energy distributions in realistic decaying systems are close to
the Lorentz distribution (\ref{328}). In these cases the decay time is determined not by the
energy dispersion (\ref{varE}), but by the energy level width $\Gamma$. 
Changing the form of the distribution function $P(E)$ at its `tail' (for $|E - E_0| \gg \Gamma$),
one can change the variance
$\Delta E$ significant1y, but the decay time will remain practically unchanged. This
fact was emphasized long ago by Fock and Krylov \cite{FoKry}. 
Therefore, to fill inequality (\ref{ETh}) with a physical content in the decay problems, 
one needs 
a more reasonable definition of energy ``uncertainty'' $\Delta E$ 
(as was questioned by Mandelstam and Tamm), 
linking it not to the energy variance (\ref{varE}), but
to some other quantity, in such a way that it would be close to the energy 
level width for distributions similar to the Lorentz one. 

Several possible definitions of this kind were proposed in Ref. \cite{183}, 
using results of 
study \cite{Bauer}, where the 
``equivalent width'' $W(\vf)$ of function $\vf(x)$  was defined as 
\be
W(\vf) =\int_{-\infty}^{\infty} \vf(x)dx/\vf(0)
\label{(346)}
\ee
(provided the integral exists and $\vf(0)\neq 0$).
If functions $f(x)$ and $\tilde{f}(y)$ are related by the Fourier transformation, 
\[
\tilde{f}(y) =\int_{-\infty}^{\infty}e^{-ixy}f(x)dx, 
\quad
f(x) =\int_{-\infty}^{\infty}e^{ixy}\tilde{f}(y) \frac{dy }{2\pi},
\]
then
\be
W(f) W( \tilde{f}) =2\pi.
\label{(347)}
\ee
The consequence of Eq. (\ref{326}) with $\hbar=1$ is the relation
\[ 
\int_{-\infty}^{\infty} e^{-i Et}\left[P(E)\right]^2\,
dE=\frac 1{2\pi}\int_{-\infty}^{\infty} \chi\left(t'\right)\chi^{*}
\left(t'- t\right)\,dt'.
\] 
Defining $f(E)=[P(E+E_0)]^2$, where $E_0$ is an arbitrary real number, one has   
\[
\tilde{f }(t)= \exp(iE_0 t)\int_{-\infty}^{\infty} \chi\left(t'+t\right)\chi^{*}\left(t'\right)\,dt'/(2\pi).
\]
\[
W(f)=\int_{-\infty}^{\infty} \left[P(E)\right]^2\, dE/[P(E_0)]^2,
\]
\[
W(\tilde{f})= \frac{\int\int dt dt' \exp(iE_0 t) \chi\left(t'+t\right)\chi^{*}\left(t'\right)}
{\int_{-\infty}^{\infty} |\chi\left(t'\right)|^2\,dt'}.
\]
Obviously
\[
|W( \tilde{f})| \le \frac{\int\int dt d\tau | \chi\left(\tau+t\right)\chi^{*}\left(\tau\right)|}
{\int_{-\infty}^{\infty} |\chi\left(\tau\right)|^2\,d\tau} =
\frac{\left[\int d\tau | \chi\left(\tau\right)|\right]^2}
{\int |\chi\left(\tau\right)|^2\,d\tau}.
\]
In addition, the identity $|W(f) W( \tilde{f})| =2\pi$ holds for any value of $E_0$ as a 
consequence of  (\ref{(347)}).
Remembering that $\chi(-t)=\chi^*(t)$ for real energy distribution function $P(E)$ and
 $Q(t) = |\chi(t)|^2$, one can arrive at the inequality
\begin{equation}
\frac {\int\left[P(E)\right]^2\,dE}{\left[P(E_0)\right]^2}\cdot\frac {\left[\int_0^{\infty}\sqrt {Q(t)}\,dt\right
]^2}{\int_0^{\infty}Q(t)\,dt}\ge\pi\hbar,
\label{350}
\end{equation}
which holds for an arbitrary value $E_0$, in particular for $E_0$ corresponding to
the maximum of function $P(E)$. Looking at the left-hand side of (\ref{350}), it seems reasonable
to introduce the following definitions of the decay time and energy uncertainty:
\begin{equation}
\tau_{*}=\frac {\left[\int_0^{\infty}\sqrt {Q(t)}\,
dt\right]^2}{4\int_0^{\infty}Q(t)\,dt}, \quad
\Delta E_{*}=\frac {\int\left[P(E)\right]^2\,dE}{\max\left
[P(E)\right]^2}.
\label{351}
\end{equation}
Then, $\tau_{*}=\tau$ for the exponential decay  (\ref{327}). 
Therefore, the lower bound for the product
$\tau_* \Delta E_*$ is the same as in (\ref{332}) or (\ref{334}), 
but with different meanings of symbols:
\begin{equation}
\tau_{*}\Delta E_{*}\ge\pi\hbar /4.
\label{353}
\end{equation}
  On the other hand, taking into account Eq. (\ref{333}), one can
define the characteristic decay time and energy uncertainty as
\begin{equation}
\tau_{**}=\frac 12\int_0^{\infty}\sqrt {Q(t)}\,dt, \quad
\Delta E_{**}=\left[\max P(E)\right]^{-1},
\label{356}
\end{equation}
 rewriting (\ref{350}) in the form 
\begin{equation}
\tau_{**}\Delta E_{**}\ge\pi\hbar /2
\label{355}
\end{equation}
(the coefficient $1/2$ in the definition of $\tau_{**}$
is chosen in order to ensure the equality $\tau_{**}=\tau_0$ for the exponential
decay law).
Three sets of decay times and energy uncertainties  are connected as follows:
\begin{equation}
\tau_{*}=\tau_{**}^2/\tau_0, \quad
\Delta E_{*}=\Delta E_{**}^2\tau_0/\pi\hbar.
\label{357}
\end{equation}

Relation (\ref{355}) becomes an equality for the exponential decay law (\ref{327}) 
with energy distribution (\ref{328}). 
The same is true for relation (\ref{353})
due to (\ref{357}). Therefore, inequalities (\ref{353}) and (\ref{355}) can be considered as
reasonable (and exact) energy-time uncertainty relations for decaying systems. 

As one can see, the replacement of the weight factor $P(E)$ by $[P(E)]^2$ 
 in the formulas for the energy ``effective variances'' enables
one to suppress the slowly decreasing ``tail'' of the Lorentz distribution function P(E) (\ref{328}). 
As a result, the ``effective variance''  proves to be of the order of the
physically acceptable width of the energy level.

\subsection{Problem of time operator}
\label{sec-oper}

Inequalities (\ref{ETh}) or (\ref{(299)}) could be derived immediately from the 
commutation relation
\be
[\hat{H}, \hat{T}] = -i\hbar,
\label{(363)}
\ee
if the time operator $\hat{T}$ existed. However, it was noticed by Pauli as far back as in 1926 
 that no Hermitian unbounded operator satisfying (\ref{(363)}) 
can exist for an {\em arbitrary\/} Hamiltonian $\hat{H}$. 
This is connected with the specific property of energy spectrum of physical systems: since the spectrum of the time operator must
be undoubtedly continuous and unbounded, the same properties must also be
inherent in the spectrum of the Hamiltonian. However, energy spectra of the majority
of physical systems are bounded from below; in addition, they can be discrete. 
Nonetheless, many people tried to find some surrogates of the time operator for various {\em specific\/}
systems or in some restricted sense. 

For the free particle Hamiltonian $\hat{H}_0=\hat{p}^2/(2m)$,
remembering the classical formula $x=pt/m + const$, one can define
a formal ``time operator'' as \cite{AharBohm61,Goto}
\be
\hat{T}_0 = m\left(\hat{p}^{-1}\hat{x} + \hat{x}\hat{p}^{-1}\right).
\label{(364)}
\ee
Although the operators $\hat{H}_0$ and $\hat{T}_0$ formally
satisfy relation (\ref{(363)}), it is clear that operator (\ref{(364)}) has a lot of defects. First of all,
it contains a singular operator $\hat{p}^{-1}$.
In addition, although operator (\ref{(364)}) in the energy representation can be reduced to the form
$\hat{T}_0 = i\hbar\partial/\partial E$, nonetheless it is non-Hermitian in the space of functions 
used in physics usually. Indeed, the hermiticity condition demands the eigenstates
$\psi(E)$ to form a complete set in the semiaxis $E> 0$, to turn into zero at $E = 0$
and to be closed with respect to the operation $\partial/\partial E$. Such a set of functions does not exist,
as was pointed out in \cite{Allcock}. Therefore, when dealing with operators like (\ref{(364)}) 
one should either ignore their unpleasant properties or define
the class of admissible wave functions in a special manner (see also \cite{Busch94,Miyamoto01} in this connection). 
The proposal to abandon the idea of unbounded time and to use instead some bounded time operators
satisfying (\ref{(363)}) was developed in \cite{Galapon02}.
 
Note that if the question about the existence of ``time operator'' is understood in a restricted
sense, e.g. as the question of finding an operator satisfying relation (\ref{(363)}) for
a given Hamiltonian, then such an operator can be found probably for
any Hamiltonian. In particular,  the prescription for how to do this
for one-dimensional systems was given in \cite{Goto81}. 
But the real problem consists in finding conditions under
which the formal ``time operator'' proves to be Hermitian (more precisely, self-conjugate).
Even more important is the possibility of physical interpretation of such an
operator as true time operator. For example, 
the problem of the energy spectrum boundedness does not exist
for the Hamiltonian $\hat{H}_F= \hat{p}^2/(2m) -F\hat{x}$ describing a particle moving in a uniform potential field. 
The spectrum of $\hat{H}_F$ is continuous and extends from $-\infty$ to $\infty$. The operator 
$\hat{T}_F = \hat{p}/F$ \cite{Busch94,Razavy67}
is Hermitian and satisfies equation (\ref{(363)}), but what is its relationship to real time? 
Moreover, $\hat{T}_F$ does not commute with the coordinate operator $\hat{x}$.
Thus, if it were the real time operator, this would mean that it is impossible
to determine simultaneously the coordinate of a particle and the time at which
this particle passes through the point with this coordinate. For these reasons
the difficulties with physical interpretation force us to treat most ``time  operators'' constructed thus far 
as purely mathematical artificial constructions without true physical meaning.
The worst feature of such ``time operators'' is that they are not universal. Instead, they should be
adjusted each time to the concrete Hamiltonian: compare operators $\hat{T}_0$ and $\hat{T}_F$.

Nonetheless, attempts to construct operators resembling time in some restricted sense continue.
One direction is to extend the Hilbert space, thus removing the problem of lower bound of the
effective Hamiltonian \cite{Rosenbaum,Bauer83}. A time-like operator for the harmonic oscillator
was constructed in \cite{Garr70}.  More general cases of systems with discrete
energy spectra were considered in Refs. \cite{Pegg98,Galapon02a,Arai08,Hall08}. 
In this connection, the entropic time--energy uncertainty relation was
introduced in \cite{Hall08}.

One may suppose that the time operator could arise naturally in the relativistic quantum mechanics:
if operators $\hat{x}_{\mu}$ exist for $\mu=1,2,3$, then operator $\hat{x}_0$ must exist as well
due to the relativistic invariance.
 Different approaches to
constructing a relativistic time operator can be found, for example, in studies
\cite{Prug,Hussar,Bunge03,Bauer14}.
However, no unambiguous results were obtained in this field.
 Perhaps the reason for this failure is that not only the time operator does not exist
in the relativistic case, but well-defined operators of the coordinates do not exist either. 
Thus, the time operator can apparently be introduced with the same degree of conventionality as the coordinate operator. 
For example, such a conventional operator was constructed by analogy with the known 
Newton--Wigner coordinate operator
\cite{NewtonWigner} in study \cite{Arshan85} (in the momentum representation for a free particle):
\be
\hat{T}_{rel} = -i\hbar\left(\frac{\partial}{\partial E} +\frac{E}{c^2 p}\frac{\partial}{\partial p}
+\frac{E}{2c^2 p^2} \right).
\label{(365)}
\ee
Then $\Delta E \Delta T \ge \hbar/2$, but operator $\hat{T}_{rel}$
 does not commute with the momentum operator:
\be
\left[\hat{T}_{rel}, \hat{\bf p}\right] = -i\hbar \frac{\hat{\bf p} E}{c^2 p^2}, \qquad
\left[\hat{T}_{rel}, \left|\hat{\bf p}\right|\right] = - \frac{i\hbar \hat{E}}{c^2 p}.
\label{(366)}
\ee
The consequence of these relations is the rigorous version of the known Landau--Peierls inequality 
\cite{LandPeierls}
\be
\Delta T \Delta p \ge \frac12 \hbar\langle v^{-1}\rangle \ge \hbar/(2c),
\label{(367)}
\ee
which relates the accuracy of the measurement of momentum to the duration of the measurement.
A small selection of other studies on the time-energy problems contains the papers
\cite{Kijowski,Vorontsov81,Braunstein96,Kobe94,Ordon01,Wang07,Olkhov07,Briggs08,HegMuga10,Denur10}.

\section{Conclusion}
The examples considered in this short mini-review represent only a small part of several hundred publications
devoted to different kinds of uncertainty relations during almost a century of studies.
Due to the lack of space, I did not touch such important areas as the measurement (``noise-disturbance'')
uncertainty relations, uncertainty relations in the quantum information theory, optics, radiophysics, 
signal analysis, thermodynamics, speed of quantum evolution, and so on.
These subjects need separate reviews.

\begin{acknowledgments}
The author acknowledges the partial support from the 
National Council for Scientific and Technological Development (CNPq)
and
Funda\c{c}\~ao de Apoio \`a Pesquisa do Distrito Federal (FAPDF), grant number 00193-00001817/2023-43.

\end{acknowledgments}

\end{document}